\documentclass[journal]{IEEEtran}
\usepackage[noadjust]{cite}
\usepackage{amsmath}
\usepackage{algorithmic}
\usepackage{array}
\usepackage{mdwmath}
\usepackage{eqparbox}
\usepackage{stfloats}
\usepackage{url}
\usepackage[pdftex]{graphicx}
\graphicspath{{../pdf/}{../jpeg/}}
\DeclareGraphicsExtensions{.pdf,.jpeg,.png}
\usepackage{amsfonts}
\usepackage[tight,footnotesize]{subfigure}
\begin{document}

\title{A Generalized Spatial Correlation Model for 3D MIMO Channels based on the Fourier Coefficients of Power Spectrums}
\author{Qurrat-Ul-Ain~Nadeem,~\IEEEmembership{Student Member,~IEEE,} Abla~Kammoun,~\IEEEmembership{Member,~IEEE,}
        M{\'e}rouane~Debbah,~\IEEEmembership{Fellow,~IEEE,} and ~ Mohamed-Slim~Alouini,~\IEEEmembership{Fellow,~IEEE}
\thanks{Manuscript received November 01, 2014; revised March 29, 2015; accepted April 16, 2015. }
\thanks{Copyright (c) 2015 IEEE. Personal use of this material is permitted. However, permission to use this material for any other purposes must be obtained from the IEEE by sending a request to pubs-permissions@ieee.org.}
\thanks{The work of Q.-U.-A. Nadeem, A. Kammoun and M. -S. Alouini was supported by a CRG 3 grant from the  Office of Sponsored Research at KAUST.  The work of M{\'e}rouane~Debbah was supported by ERC Starting Grant 305123 MORE (Advanced Mathematical Tools for Complex Network Engineering).}   
\thanks{Q.-U.-A. Nadeem, A. Kammoun and M.-S. Alouini are with the Computer, Electrical and Mathematical Sciences and Engineering (CEMSE) Division, King Abdullah University of Science and Technology (KAUST), Thuwal, Makkah Province, Saudi Arabia 23955-6900 (e-mail: \{qurratulain.nadeem,abla.kammoun,slim.alouini\}@kaust.edu.sa)}
\thanks{M. Debbah is  with  Sup{\'e}lec, Gif-sur-Yvette, France and Mathematical and Algorithmic Sciences Lab, Huawei France R\&D, Paris, France (e-mail: merouane.debbah@huawei.com, merouane.debbah@supelec.fr).}
}

\markboth{IEEE TRANSACTIONS ON SIGNAL PROCESSING,~Vol.~xx, No.~x, xx~2014}%
{Shell \MakeLowercase{\textit{et al.}}: Bare Demo of IEEEtran.cls for Journals}

\maketitle

\begin{abstract}
Previous studies have confirmed the adverse impact of fading correlation on the mutual information (MI) of two-dimensional (2D) multiple-input multiple-output (MIMO) systems. More recently, the trend is to enhance the system performance by exploiting the channel's degrees of freedom in the elevation, which necessitates the derivation and characterization of three-dimensional (3D) channels in the presence of spatial correlation. In this paper, an exact closed-form expression for the Spatial Correlation Function (SCF) is derived for 3D MIMO channels. This novel SCF is developed for a uniform linear array of antennas with nonisotropic antenna patterns. The proposed method resorts to the spherical harmonic expansion (SHE) of plane waves and the trigonometric expansion of Legendre and associated Legendre polynomials. The resulting expression depends on the underlying arbitrary angular distributions and antenna patterns through the Fourier Series (FS) coefficients of power azimuth and elevation spectrums. The novelty of the proposed method lies in the SCF being valid for any 3D propagation environment. The developed SCF  determines the covariance matrices at the transmitter and the receiver that form the Kronecker channel model. In order to quantify the effects of correlation on the system performance, the information-theoretic deterministic equivalents of the MI for the Kronecker model are utilized in both mono-user and multi-user cases. Numerical results validate the proposed analytical expressions and elucidate the dependence of the system performance on azimuth and elevation angular spreads and antenna patterns. Some useful insights into the behaviour of MI as a function of downtilt angles are provided. The derived model will help evaluate the performance of correlated 3D MIMO channels in the future.  
\end{abstract}

\begin{IEEEkeywords}
3D multiple-input multiple-output (MIMO) systems,  spatial correlation, power azimuth spectrum, power elevation spectrum, elevation beamforming, mutual information.
\end{IEEEkeywords}


\section{Introduction}

Multiple-input multiple-output (MIMO) systems have remained a subject of interest in wireless communications over the past decade due to the significant gains they offer in terms of spectral efficiency by exploiting the multipath richness of the channel. The increased spatial degrees of freedom not only provide diversity and interference cancellation gains but also help achieve a significant multiplexing gain by opening several parallel sub-channels. Pioneer work in this area by Telatar \cite{Telatar} and Foschini \cite{foschini} realized that capacity can potentially scale linearly with the minimum number of transmit (Tx) and receive (Rx) antennas for channel matrices with centered, independent and identically distributed (i.i.d) elements. These MIMO systems were designed to support antenna configurations capable of adaptation in the azimuth only. However recent measurement campaigns demonstrated that elevation has a significant impact on the system performance \cite{elevationcamp}, \cite{elevationcamp1}.  Exploiting the channel's degrees of freedom in the elevation can further enhance the system performance by benefiting from the richness of real channels. This has recently become a subject of interest  among researchers and industrials. The reason can be attributed to its potential to open up possibilities for a variety of strategies like user specific elevation beamforming and cell-splitting. Encouraged by the initial implementations of this technology \cite{practicals}, the 3GPP is now working on defining future mobile communication standards in the frame of the study items on three-dimensional (3D) beamforming \cite{3GPP3D}. 

The discussion on the conspicuous advantages of 3D MIMO systems must be amalgamated with the observation that it is the orthogonality of the subchannels constituting the MIMO system that determines the extent of the multiplexing gain that can be realized. Large capacity gains can only be realized when the subchannels are potentially decorrelated. However in realistic propagation environments, the promised theoretical gains are not realized due to the significant spatial correlation present in the MIMO channel \cite{kronecker, kronecker_model, correlation5, kronecker_model2, correlationpresence3, correlationpresence, correlationpresence2}. Therefore assuming the channel coefficients to be i.i.d is an oversimplification of the problems encountered in realistic propagation environments. 

The need to investigate the impact of spatial correlation on the performance of MIMO systems is acknowledged and well-known among researchers. However, most of the spatial correlation models proposed in literature are developed for 2D channel models and ignore the elevation \cite{correlation5, additional_correlation, correlation2, correlation3,correlation7,correlation9}. These 2D models assume an omnidirectional radiation of energy in the elevation. However, with the advent of smart antennas \cite{smart_antenna}, it is important to take into account the characteristics and patterns of the directional antennas that are being widely deployed now.  These antenna patterns have often been described in literature using spherical harmonics \cite{pattern1, pattern2}, but this has not been done in the context of spatial correlation.  A recent approach to incorporate these patterns in the correlation function was proposed in \cite{correlation1}, but the method could not admit a closed-form solution for this case and succumbed to numerical integration methods. 

The existing correlation models  are derived for a particular distribution of the Angle of Departure (AoD) and Angle of Arrival (AoA)  such as uniform, Gaussian, Von Mises, cosine or Laplacian \cite{correlation5,additional_correlation,correlation3, Shafi06polarizedmimo, correlation1, correlation6,correlation4,correlation2,correlation7,correlation9}.  In \cite{correlation3}, approximate closed-form expressions for the spatial correlation coefficients for clustered MIMO channel models were derived for Laplacian azimuth AoA distribution. The proposed method makes small angle spread approximation for uniform linear and circular arrays and offers significant gains in terms of computational cost. Such assumptions on angular distributions can lead to useful closed-form expressions but do not accurately represent the characteristics of realistic propagation environments.  In \cite{correlation2}, the authors derived exact closed-form expressions for the spatial correlation between Rx antenna elements for cosine, Gaussian and Von Mises azimuth AoA distributions. The use of Von Mises was shown to simplify the expressions and the impact of mutual coupling on the correlation was studied.  

The notion of spatial correlation in 3D propagation environments has been addressed in some research works. An important contribution in this area appears in \cite{Shafi06polarizedmimo}. The authors developed closed-form expressions for the spatial correlation and large system ergodic mutual information (MI) for a 3D cross-polarized channel model, assuming the angles to be distributed according to Von Mises distribution. The authors in \cite{correlation1}, showed that elevation plays a crucial role in determining the Spatial Correlation Function (SCF). The derivation is based on the spherical harmonic expansion (SHE) of plane waves and assumes the distribution of AoA to be 3D Von Mises-Fisher. In \cite{correlation4}, closed-form expressions for spatial fading correlation functions of several omnidirectional antenna arrays in a 3D MIMO channel were derived and used for the evaluation of channel capacity. The derived results were expressed as a function of angular and array parameters and used to study the impact of azimuth and elevation angular spreads on the MI. However, this work assumes the angular distributions to be uniform. The analysis in \cite{general} uses SHE of plane waves to compute the closed-form expressions for the correlation that can be applied to a variety of angular distributions. Although the tools presented are handy, the proposed closed-form solutions require certain assumptions to be made on the propagation environment. Even the simple assumption that the angles are uniformly distributed resulted in integrals involving Legendre polynomials that could not be expressed in a closed-form. Such assumptions neither aptly represent the characteristics of  realistic propagation environments nor make the proposed method truly generic in nature. 

In order to quantify the effect of correlation, it is important to derive and simulate the correlated MIMO channels and characterize the information-theoretic MI for them. There are two widely used approaches to model these channels \cite{correlation3}. The first one is the parametric approach, in which the propagation paths are described using statistical parameters without being physically positioned. Channel realizations are generated by summing the contributions of multiple paths  (plane waves), with specific channel parameters like delay, amplitude, AoA and AoD.  The second approach is nonparametric, wherein the SCF is used to determine the covariance matrices at the transmitter and receiver. These matrices are then employed to reproduce the spatial correlation across the MIMO channel. An example is the Kronecker channel model, which is useful for the evaluation of theoretical MI.  In this context, well-known results from Random Matrix Theory (RMT) have been employed to characterize the distribution of the MI of these channels in the asymptotic regime as the number of antennas at the base station (BS) and mobile station (MS) tend to grow large \cite{walid_mutual_information,walid_mutual_information1,deterministic}, \cite{kronecker_model}. Such theoretical results enable better understanding of the impact of the correlation on the MI. These so-called deterministic equivalents are reasonably tight even at moderate values of the number of antennas.  

The aims of this paper are fourfold. First is to develop an exact closed-form expression for the SCF for 3D MIMO channels that can be used for any arbitrary choice of  antenna patterns and distribution of azimuth and elevation AoD and AoA. The parametric 3D channel model used in the derivation is inspired from the models presented in standards like 3GPP SCM \cite{SCM}, WINNER+ \cite{Winner+} and ITU \cite{ITU}. To get an analytically tractable closed-form solution, the SHE of plane waves and properties of Legendre and associated Legendre polynomials are exploited. The final expressions for the SCF developed for a uniform linear array of antennas are presented in \textit{Theorem 1} [(\ref{theorem}), (\ref{tcorr}), (\ref{rcorr})]. The closed-form expressions depend on the underlying arbitrary angular distributions and antenna patterns through the Fourier Series (FS) coefficients of Power Azimuth Spectrum (PAS) and Power Elevation Spectrum (PES). To the best of authors' knowledge, a SCF that works for the 3D channel model without making any assumptions on the underlying angular distributions and antenna patterns has not been developed before. The second aim of this work is to validate the proposed SCF via simulations for angular distributions and antenna patterns specified in the standards. The FS coefficients are computed and used to obtain the correlation coefficients that coincide with the Monte-Carlo simulated results. The third aim is to use the nonparametric Kronecker channel model for the evaluation of MI in the mono-user case. The developed SCF is used to determine the covariance matrices at the transmitter and receiver that form the Kronecker model. The pinhole phenomenon is discussed and illustrated as a restriction to the nonparametric Kronecker channel model as compared to the parametric channel model. The theoretical analysis for the mono-user case makes use of the deterministic equivalent of the MI presented in \cite{walid_mutual_information} and studies the effect of angular parameters parametrized by azimuth and elevation angular spreads on the MI.  An interesting interplay between vertical antenna pattern and elevation spread is observed. The final goal of this work is to provide a flavor of the performance gains realizable through the meticulous selection of the transmit antenna downtilt angles in a multi-user scenario. The MI analysis makes use of the deterministic equivalent of the signal-to-interference plus noise ratio (SINR) in \cite{SINRdeterministic} with regularized zero forcing (RZF) precoding at the BS to mitigate inter-user interference. The researchers and industrials interested in using our correlation model need to provide only the FS coefficients of the PAS and PES they are using for the evaluation of their work. 

\begin{table}[t]
\normalsize
\caption{List of important symbols used in sections II-IV.}
\centering
\begin{tabular}{|c|c|}
\hline
Symbol & Description \\ 
\hline
$\theta$, $\vartheta$ & Elevation AoD and AoA respectively.  \\ 
 $\phi$, $\varphi$ & Azimuth AoD and AoA respectively. \\ 
 $N$ & Number of propagation paths.  \\
 $\alpha_{n}$ & Complex amplitude of the $n^{th}$ path.  \\ 
 $\theta_{tilt}$ & Elevation angle of antenna boresight.  \\ 
$g_{t}(\phi,\theta,\theta_{tilt})$ & Tx antenna pattern. \\
$g_{r}(\varphi,\vartheta)$ & Rx antenna pattern. \\
$g_{t,H}(\phi)$ & Horizontal antenna pattern. \\ 
$g_{t,V}(\theta,\theta_{tilt})$ & Vertical antenna pattern. \\
 $\textbf{k}_{t}$ & Transmitted wave vector.  \\
$\textbf{k}_{r}$ & Received wave vector. \\
$\textbf{x}$ & Location vector of an antenna in $\mathbb{R}^{3}$. \\ 
  $\hat{\textbf{v}}$ & Direction of wave propagation. \\ 
	$\phi_{3dB}$ & Horizontal 3 $\rm{dB}$ beamwidth. \\ 
	$\theta_{3dB}$ & Vertical 3 $\rm{dB}$ beamwidth. \\ 
$G_{p,max}$ & Antenna gain. \\
$N_{BS}$ & Number of Tx antennas. \\
$N_{MS}$ & Number of Rx antennas. \\
$d_{t}$ & Distance between Tx antennas. \\
$d_{r}$ & Distance between Rx antennas. \\
$PAS$ & Power azimuth spectrum. \\
$PES$ & Power elevation spectrum. \\
$\theta_{0}$ & Mean AoD in the elevation. \\
$\sigma$ & Angular spread in the elevation. \\
$\rho_{t}(s-s')$ & Correlation between Tx antennas $s$, $s'$. \\
$\rho_{r}(u-u')$ & Correlation between Rx antennas $u$, $u'$. \\
$j_{n}$ & Bessel function of order $n$. \\
$P_{n}$ & Legendre polynomial function of order $n$. \\
$P_{n}^{m}$ & Associated Legendre polynomials. \\
$\bar{P}_{n}^{m}$ & $\sqrt{(n+\frac{1}{2})\frac{(n-m)!}{(n+m)!}}P_{n}^{m}(\text{x})$. \\
$\beta_{t}$, $\beta_{r}$ & $\frac{2\pi}{\lambda}d_{t}$ and $\frac{2\pi}{\lambda}d_{r}$ respectively. \\
$p_{n},c_{2n,2k}^{2m}$ &  Legendre coefficients for even orders. \\
$d_{2n-1,2k-1}^{2m-1}$ & Legendre coefficients for odd orders. \\
$a_{\phi},b_{\phi}$ &  FS coefficients of PAS. \\
$a_{\theta},b_{\theta}$ &  FS coefficients of PES. \\
$N_{0}$ & Number of terms summed over $n$  \\
& in the proposed SCF. \\
$\mu$ & Mean AoD in the azimuth. \\
$\kappa$ & Angular spread in the azimuth. \\
\hline
\end{tabular}
\label{table1}
\end{table}

This paper is organized as follows. The 3D channel model, antenna configuration, PAS and PES are explained in Section II.  In section III, we present an analytical derivation of the proposed closed-form expression for the generalized SCF. In Section IV, we provide simulation results that validate the developed SCF, adhering to most of the guidelines provided in the standards. In section V, we present the nonparametric Kronecker channel model and recall well-known results on the deterministic equivalents of the MI of this model in the mono-user and multi-user systems. The performance of these systems is investigated as a function of channel and array parameters through numerical results. Finally in section VI, some concluding remarks are drawn.


\section{Channel Model and Power Spectrums}

Prior to proceeding into the derivation of the SCF for MIMO channels, it is vital to explain the characteristics of the 3D channel model and antenna configuration under investigation. MIMO systems of current LTE releases do not support antenna configurations capable of adaption in the elevation. However, encouraged by the potential of elevation beamforming to enhance system performance, some standardized channel models have started to emerge that define the next generation 3D channels. We base the evaluation of our work on these channel models after making some realistic assumptions on the channel parameters. To improve the clarity of mathematical exposition, the important symbols used in sections II-IV are listed in Table \ref{table1}. 

\subsection{Standardized 3D Channel Model}
The MIMO channel model for which the SCF is derived is inspired from the standardized models like 3GPP SCM \cite{SCM}, ITU \cite{ITU} and WINNER \cite{Winner}. These standards follow a system level, stochastic channel modeling approach wherein, the propagation paths are described using statistical parameters without being physically positioned. Channel realizations are generated by summing the contributions of multiple paths with specific parameters like delay, amplitude, AoA and AoD.  

\begin{figure}[!b]
\centering
\includegraphics[width=3 in]{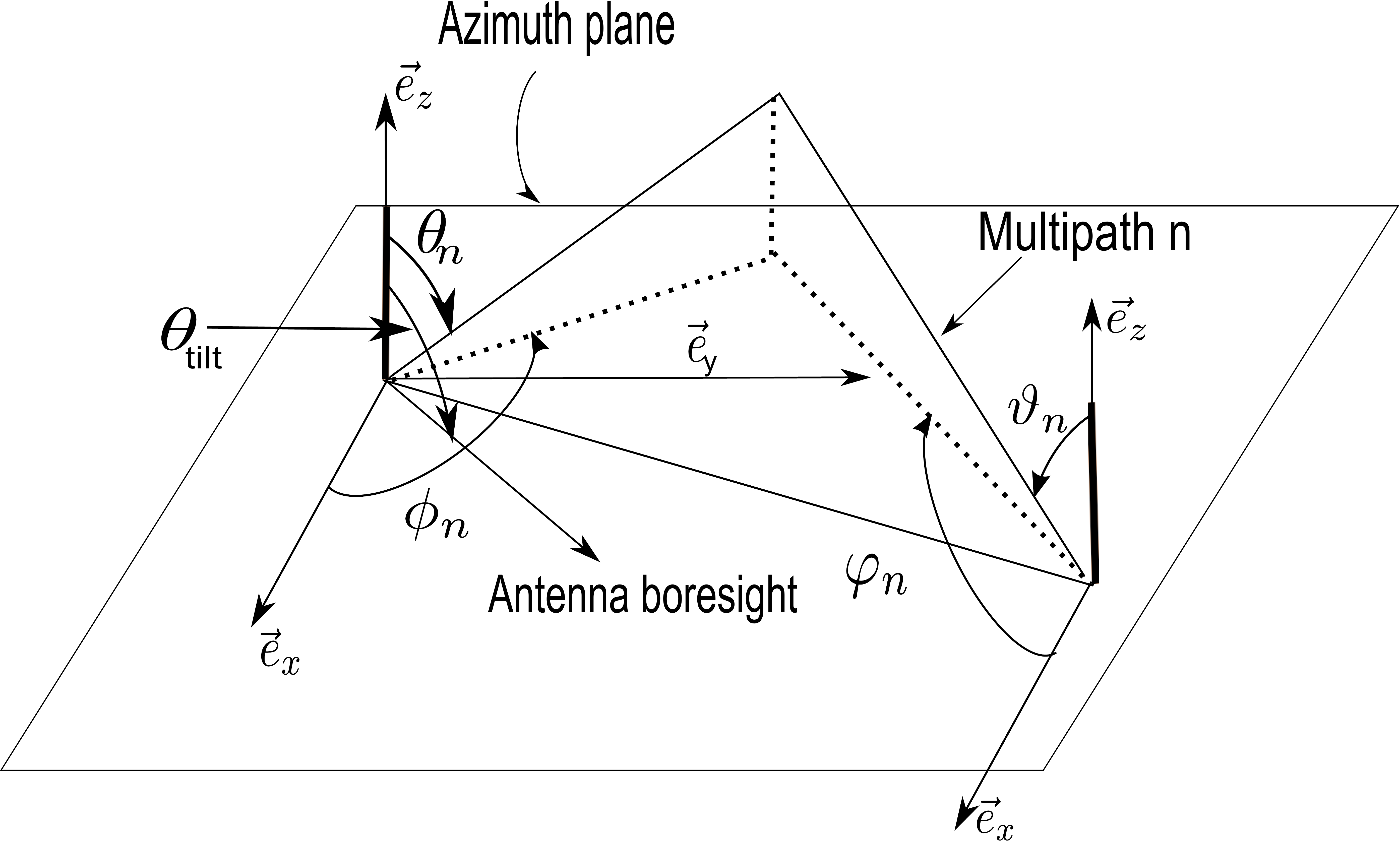}
\caption{3D channel model.}
\label{channelfig}
\end{figure}

\begin{figure*}[!t]
\normalsize
\setcounter{equation}{7}
\begin{align}
\label{channel}
& [\textbf{H}]_{su}= \sum\limits_{n=1}^N  \alpha_{n} \sqrt{g_{t}(\phi_{n},\theta_{n},\theta_{tilt})} \exp \left(i k (s-1) d_{t} \sin\phi_{n}\sin\theta_{n}\right) \sqrt{g_{r}(\varphi_{n},\vartheta_{n})} \exp\left(i k  (u-1) d_{r}\sin\varphi_{n}\sin\vartheta_{n}\right), 
\end{align}
\hrulefill
\vspace*{4pt}
\end{figure*}

Almost all system level based standards are 2D. However, owing to the growing interest in 3D beamforming, extensions of these standards to the 3D case have started to emerge recently in \cite{3GPP3D}, \cite{Winner+}.  Based on the aforementioned standards, the effective channel between the BS antenna $s$ and the MS antenna $u$ is given by \cite{Winner+}, \cite{drabla},
\setcounter{equation}{0}
\begin{align}
\label{channel1}
\text{[\textbf{H}]}_{su}&=\sum\limits_{n=1}^N  \alpha_{n} \sqrt{g_{t}(\phi_{n},\theta_{n},\theta_{tilt})}  \sqrt{g_{r}(\varphi_{n},\vartheta_{n})} [\textbf{a}_{r}(\varphi_{n}, \vartheta_{n})]_{u} \nonumber \\
& \times [\textbf{a}_{t}(\phi_{n}, \theta_{n})]_{s},
\end{align}
where  $\phi_{n}$ and $\theta_{n}$ are the azimuth and elevation AoD of the $n^{th}$ path respectively, $\varphi_{n}$ and $\vartheta_{n}$ are the azimuth and elevation AoA of the $n^{th}$ path respectively and $\theta_{tilt}$ is the elevation angle of the antenna boresight. Note that $\theta_{tilt}=90^{o}$ corresponds to zero electrical downtilt.  ${\alpha}_{n}$ is the complex amplitude of the $n^{th}$ path. The complex amplitudes are assumed to be i.i.d zero mean, $\frac{1}{N}$ variance Gaussian RVs. Also $\sqrt{\text{g}_{t}(\phi_{n},\theta_{n},\theta_{tilt})}$ and $\sqrt{\text{g}_{r}(\varphi_{n},\vartheta_{n})}$ are the global patterns of Tx and Rx antennas respectively  where $g_{t}(\phi_{n},\theta_{n},\theta_{tilt}) \approx g_{t,H}(\phi_{n})g_{t,V}(\theta_{n},\theta_{tilt})$ and $g_{r}(\varphi_{n},\vartheta_{n}) \approx g_{r,H}(\varphi_{n})g_{r,V}(\vartheta_{n})$. Note that $g_{t,H}(\phi), g_{r,H}(\varphi)$ are the horizontal antenna patterns and $g_{t,V}(\theta,\theta_{tilt}), g_{r,V}(\vartheta)$ are the vertical antenna patterns. Moreover, vectors $\textbf{a}_{t}(\phi, \theta)$ and $\textbf{a}_{r}(\varphi, \vartheta)$ are the array responses of the Tx and Rx antennas respectively whose entries are given by,
\begin{align}
[\textbf{a}_{t}(\phi, \theta)]_{s}&=\exp(i \textbf{k}_{t}\textbf{.}\textbf{x}_{s}), \\
[\textbf{a}_{r}(\varphi, \vartheta)]_{u}&=\exp(i \textbf{k}_{r}\textbf{.}\textbf{x}_{u}),
\end{align}
where \textbf{.} is the scalar product, $\textbf{x}_{s}$ and $\textbf{x}_{u}$ are the location vectors of the $s^{th}$ Tx antenna and the $u^{th}$ Rx antenna respectively, $\textbf{k}_{t}$ and $\textbf{k}_{r}$ are the Tx and Rx wave vectors respectively, where $\textbf{k}=\frac{2\pi}{\lambda}\hat{\textbf{v}}$, with $\lambda$ being the carrier wavelength and $\hat{\textbf{v}}$ being the direction of wave propagation. Fig. \ref{channelfig} illustrates the 3D channel model being considered. It is evident that $\theta,\theta_{tilt}, \vartheta \in (0, \pi)$ and $\phi,\varphi \in (-\pi, \pi)$.  

\subsection{Antenna Configuration}
\begin{figure}[!b]
\centering
\includegraphics[width=2 in]{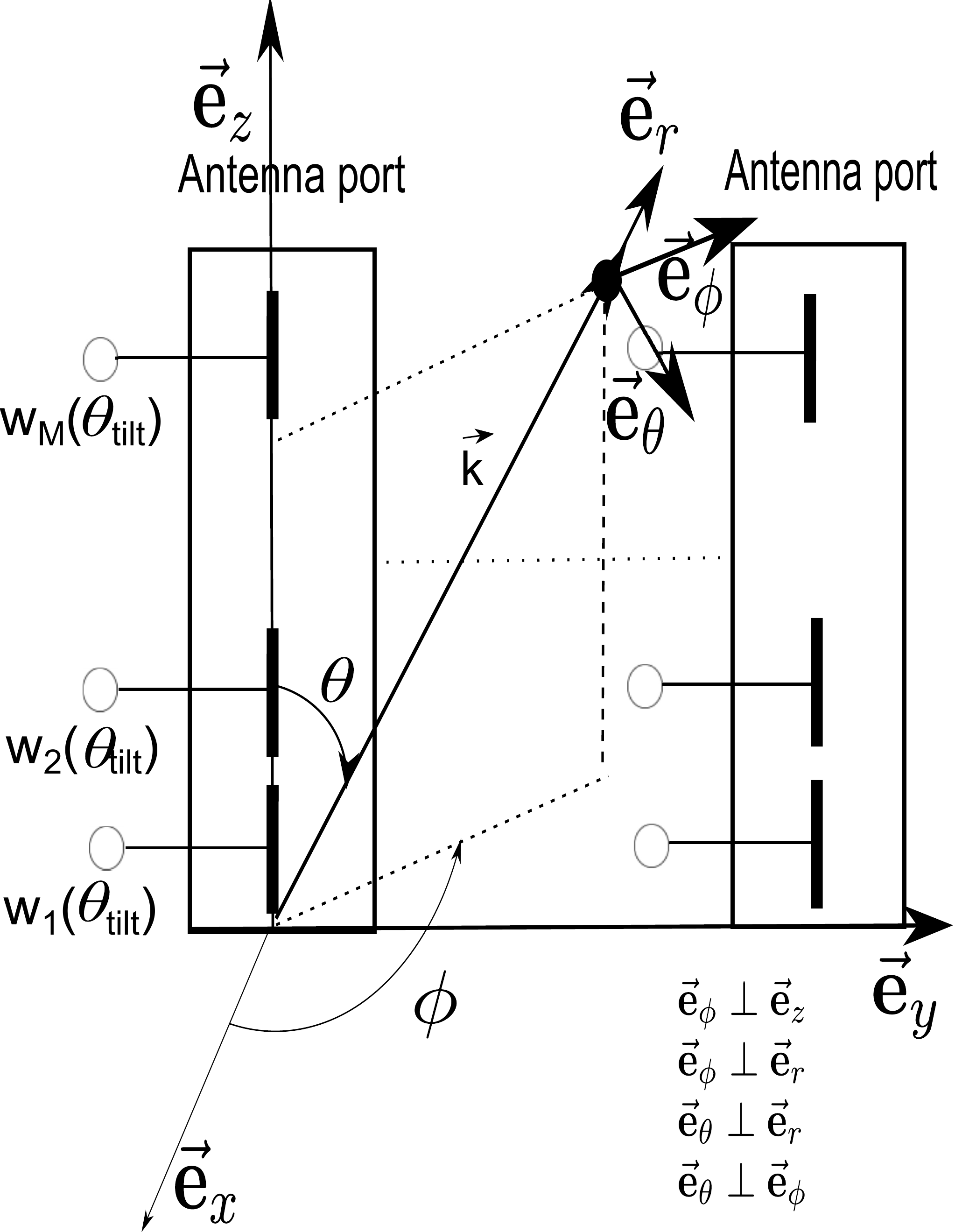}
\caption{Antenna configuration.}
\label{antenna}
\end{figure}

If $(\hat{\textbf{e}}_{r}, \hat{\textbf{e}}_{\theta}, \hat{\textbf{e}}_{\phi})$ is the spherical coordinate system, then vertical polarization refers to the polarization along $\hat{\textbf{e}}_{\theta}$ and horizontal polarization refers to polarization along $\hat{\textbf{e}}_{\phi}$. Each antenna port comprises of vertically stacked antenna elements that determine the effective antenna port pattern. For the purpose of this work, vertically polarized antenna elements are considered. The antenna ports are placed at fixed positions along $\hat{\textbf{e}}_{y}$, with the elements in each port aligned along $\hat{\textbf{e}}_{z}$ as shown in Fig. \ref{antenna}.  There are $N_{BS}$ and $N_{MS}$ antenna ports at the BS and MS respectively. The same Tx signal is fed to all elements in a port with corresponding weights to achieve the desired directivity. The MS sees each antenna port as a single antenna because all elements carry the same signal. Therefore, we are interested in the channel between the Tx antenna port and the Rx antenna port. 

In theory, the global  antenna pattern of a port depends on the patterns of the elements within it and their corresponding weights. However to enable an abstraction of the role played by these  elements to perform elevation beamforming, standards like 3GPP and ITU approximate the pattern of each antenna port by a narrow beam in the elevation. The combined pattern considered in $\rm{dB}$ is as follows \cite{ITU}, \cite{drabla},
\setcounter{equation}{3}
\begin{align}
\label{port}
A_{p}(\phi,\theta,\theta_{tilt})&= G_{p,max}-\text{min}\{-(A_{H}(\phi)+A_{V}(\theta,\theta_{tilt})),20\},
\end{align}
where,
\begin{align}
\label{minpattern}
A_{H}(\phi)&= -\text{min}\left[ 12 \left(\frac{\phi}{\phi_{3dB}}\right)^2, 20 \right] \rm{dB}, \nonumber \\
A_{V}(\theta,\theta_{tilt})&= -\text{min} \left[12 \left(\frac{\theta-\theta_{tilt}}{\theta_{3dB}}\right)^2,20 \right] \rm{dB}.
\end{align}
Therefore the horizontal and vertical antenna patterns at the transmitter can be approximated as,
\begin{align}
\label{hpattern}
g_{t,H}(\phi)&= - 12\left(\frac{\phi}{\phi_{3dB}}\right)^2 \rm{dB}, \\
\label{vpattern}
g_{t,V}(\theta,\theta_{tilt})&= -12 \left(\frac{\theta-\theta_{tilt}}{\theta_{3dB}}\right)^2 \rm{dB}.
\end{align}

$G_{p,max}=17$ $\rm{dBi}$, $\phi_{3dB}$ is the horizontal 3 $\rm{dB}$ beamwidth and $\theta_{3dB}$ is the vertical 3 $\rm{dB}$ beamwidth. The individual antenna radiation pattern at the MS, $g_{r}(\varphi,\vartheta)$, is taken to be $0$ $\rm{dB}$ in the standards. It is important to note here that we provided the expressions for the antenna patterns proposed by the standards only for the sake of completeness of the standardized 3D channel model. Also, they will be used to validate our proposed SCF later in section IV. However, the development of the SCF provided in this work is independent of the form of the underlying antenna patterns as long as $g_{t}(\phi_{n},\theta_{n},\theta_{tilt}) \approx g_{t,H}(\phi_{n})g_{t,V}(\theta_{n},\theta_{tilt})$ and $g_{r}(\varphi_{n},\vartheta_{n}) \approx g_{r,H}(\varphi_{n})g_{r,V}(\vartheta_{n})$. 

Given the antenna configuration shown in Fig. \ref{antenna}, $\hat{\textbf{v}}_{t}\textbf{.}\hat{\textbf{x}}_{s}=\sin\phi_{n}\sin\theta_{n}$ and $\hat{\textbf{v}}_{r}\textbf{.}\hat{\textbf{x}}_{u}=\sin\varphi_{n}\sin\vartheta_{n}$. The effective radio channel given in (\ref{channel1}) can hence be written as (\ref{channel}), where $s=1, \dots N_{BS}$ and $u=1, \dots N_{MS}$.

\subsection{Power Azimuth and Elevation Spectrum}
Power azimuth spectrum (PAS) and power elevation spectrum (PES) are important statistical properties of wireless channels and are shown to play an important role in determining the spatial correlation present in the MIMO channel. They provide a measure of the power distribution upon the azimuth AoD and AoA and the elevation AoD and AoA respectively. 

Observing that the integral of the product of  angular power density function of the azimuth AoD/AoA and the horizontal antenna pattern at the BS/MS yields the expected power transmitted/received by the directional antenna in the azimuth \cite{PAS}, we define PAS at the transmitter as follows,
\setcounter{equation}{8}
\begin{align}
\label{PAS}
\text{PAS}_{t}(\phi)&= g_{t,H}(\phi) p_{\phi}(\phi),
\end{align}
where the angular power density function $p_{\phi}(\phi)$=$f_{\phi}(\phi)$, the probability density function of azimuth angle. Therefore,
\begin{align}
\label{condPAS}
\int_{-\pi}^{\pi} p_{\phi}(\phi) d\phi &=1.
\end{align}
Similarly PES at the Tx side is defined as,
\begin{align}
\label{PES}
\text{PES}_{t}(\theta,\theta_{tilt}) &= g_{t,V}(\theta,\theta_{tilt}) p_{\theta}(\theta).
\end{align}
The elevation angular power density function, $p_{\theta}(\theta)$=$\frac{f_{\theta}(\theta)}{\sin(\theta)}$, which implies \cite{Kalliola02angularpower},
\begin{align}
\label{condPES}
&\int_{0}^{2\pi}  p_{\theta}(\theta)\sin(\theta) d\theta =1.
\end{align}
The same definitions and conditions can be extended to $\text{PAS}_{r}$ and $\text{PES}_{r}$. Note that the limits taken in (\ref{condPES}) are (0, 2$\pi$) instead of (0, $\pi$), which is the range over which $\theta$ is defined. This extension in limits, which would later assist in expressing SCF in terms of the FS coefficients of PES, entails that we define $f_{\theta}(\theta)$ to be zero from $\pi$ to $2\pi$. This is generally true because the elevation angular density spectrums used in standards decay exponentially with $\theta$.

\textit{Example}:
The PES can be well fitted by the Laplace distribution \cite{PES1}. The elevation angles are thus generated using,
\setcounter{equation}{12} 
\begin{align}
\label{laplace}
f_{\theta}(\theta) \propto \exp\left(-\frac{\sqrt{2}|\theta-\theta_{0}|}{\sigma}\right)\sin{\theta},
\end{align}
where $\sigma$ is the spread in the elevation direction and $\theta_{0}$ is the mean AoD in the elevation. The density function decays exponentially and is zero for $\theta \notin [0,\pi].$ Hence to determine the constant of proportionality $A$, we can use condition (\ref{condPES}) and the observation that $f_{\theta}(\theta)=0$ for $\theta \notin [0,\pi]$, 
\begin{align}
&\int_{0}^{\pi} A \exp\left(-\frac{\sqrt{2}|\theta-\theta_{0}|}{\sigma}\right)\sin\theta d\theta=1, \\
&A=\frac{2+\sigma^{2}}{2\sqrt{2}\sigma \sin\theta_{0}+2\sigma^{2}e^{-\frac{\pi}{\sqrt{2}\sigma}}\cosh\left(\frac{\sqrt{2}(\frac{\pi}{2}-\theta_{0})}{\sigma}\right)}. 
\end{align}

\begin{figure*}[!t]
\normalsize
\setcounter{equation}{20}
\begin{align}
\label{legend}
&P_{n}(\cos \gamma)=P_{n}(\cos\theta_{1})P_{n}(\cos\theta_{2})+2\sum_{m=1}^{n} \frac{(n-m)!}{(n+m)!}P_{n}^{m} (\cos\theta_{1})P_{n}^{m} (\cos\theta_{2}) \cos[m(\phi_{1}-\phi_{2})] \\
\label{exp}
&\exp\left(i\frac{2\pi}{\lambda}d_{t}(s-s')\sin\phi\sin\theta\right) = \sum_{n=0}^{\infty} i^{n}(2n+1)j_{n}\left(\frac{2\pi}{\lambda}d_{t}|s-s'|\right) P_{n}(\sin\phi\sin\theta) \\
\label{int}
&\rho_{t}(s-s')=\mathbb{E}\Big[g_{t}(\phi,\theta,\theta_{tilt}) \sum_{n=0}^{\infty} i^{n}(2n+1)j_{n}\left(\frac{2\pi}{\lambda}d_{t}|s-s'|\right) \Big( P_{n}(\cos\theta)P_{n}(0)+2\sum_{m=1}^{n} \frac{(n-m)!}{(n+m)!} \nonumber \\
& \times P_{n}^{m} (\cos\theta)P_{n}^{m} (0) \cos\left(m\left(\phi-\frac{\pi}{2}\right)\right)   \Big)\Big]
\end{align}
\setcounter{equation}{23}
\begin{align}
\label{prop}
&\rho_{t}(s-s')=\mathbb{E}[g_{t}(\phi,\theta,\theta_{tilt}) ]j_{0}\left(\beta_{t}|s-s'|\right)  + \sum_{n=1}^{\infty} (-1)^{n}(4n+1)j_{2n}\left(\beta_{t}|s-s'|\right){P}_{2n}(0)\mathbb{E}[{P}_{2n}(\cos\theta)g_{t,V}(\theta,\theta_{tilt})]\mathbb{E}[g_{t,H}(\phi)] \nonumber \\  
&+\sum_{n=1}^{\infty}4 (-1)^{n} j_{2n}\left(\beta_{t}|s-s'|\right)  \left(\sum_{m=1}^{n} (-1)^{m}  \bar{P}_{2n}^{2m} (0)  \mathbb{E}[\bar{P}_{2n}^{2m} (\cos\theta)g_{t,V}(\theta,\theta_{tilt})] \mathbb{E}[\cos(2m\phi) g_{t,H}(\phi)]  \right)    \\  
&+\sum_{n=1}^{\infty}4 i (-1)^{n} j_{2n-1}\left(\beta_{t}|s-s'|\right)  \Big(\sum_{m=1}^{n} (-1)^{m}  \bar{P}_{2n-1}^{2m-1} (0) \mathbb{E}[\bar{P}_{2n-1}^{2m-1} (\cos\theta) g_{t,V}(\theta,\theta_{tilt})]\mathbb{E}[\sin((2m-1)\phi)g_{t,H}(\phi)]   \Big) \nonumber
\end{align}
\hrulefill
\end{figure*} 


\section{Wavefield Decomposition and Spatial Correlation Function}

In this section, we rigorously derive a generic analytical expression for the SCF considering realistic antenna patterns and arbitrary distributions of AoDs and AoAs. 

Note that since the parameters describing the multipaths are i.i.d, the double sum in $\mathbb{E}[\textbf{H}_{su}\textbf{H}_{s'u'}^{H}]$ can be simplified to $N$ multiplied by $\mathbb{E}[|\alpha|^{2}]$ and the terms defined in (\ref{Tx}) and (\ref{Rx}) below. Therefore, it can be seen directly from (\ref{channel}) that for i.i.d zero mean $\alpha$'s with $\mathbb{E}[|\alpha|^{2}]=\frac{1}{N}$, the spatial correlation between the channels constituted by any pair of Tx and Rx antenna ports can be expressed as a product of the correlation between Tx antenna ports and the correlation between Rx antenna ports, i.e.,
\setcounter{equation}{15}
\begin{align}
\label{SCF}
\text{SCF} &= \mathbb{E}[[\textbf{H}_{su} \textbf{H}_{s'u'}^{H}] =\rho_{t}(s-s') \rho_{r}(u-u'),
\end{align} 
where,
\begin{align}
\label{Tx}
\rho_{t}(s-s')&=\mathbb{E}\left[g_{t}(\phi,\theta, \theta_{tilt})\exp\left(i\frac{2\pi}{\lambda}d_{t}(s-s')\sin\phi\sin\theta\right)\right] \\
\label{Rx}
\rho_{r}(u-u')&=\mathbb{E}\left[g_{r}(\varphi,\vartheta)\exp\left(i\frac{2\pi}{\lambda}d_{r}(u-u')\sin\varphi\sin\vartheta\right)\right]
\end{align}

We derive the closed-form expression for the correlation between Tx antennas given in (\ref{Tx}). This can be extended to the correlation between Rx antennas and the product of the two would yield the SCF between the channels.

\subsection{Spherical Harmonic Expansion of Plane Waves}
In a 3D propagation environment, the array responses of Tx and Rx antennas can be expanded using spherical decomposition for plane waves. Using the Jacobi-Anger expansion, a plane electromagnetic wave can be expressed as a superposition of spherical waves \cite{jacobi},
\setcounter{equation}{18}
\begin{align}
\label{Jacobi}
& e^{ik\textbf{x}.\hat{\textbf{v}}} =\sum_{n=0}^{\infty} i^{n}(2n+1)j_{n}(k||\textbf{x}||) P_{n}\left(\hat{\textbf{x}}.\hat{\textbf{v}}\right), \textbf{x}\in \mathbb{R}^{3},
\end{align}
where $k$=$\frac{2\pi}{\lambda}$ is the wave number, $\hat{\textbf{v}}$ is a unit vector in the direction of wave propagation,  $\textbf{x}$ is the location vector of the antenna in $\mathbb{R}^{3}$, $j_{n}$ is the spherical Bessel function of order $n$ and $P_{n}$ is the Legendre polynomial function of order $n$. 

We also state here the Legendre addition theorem of spherical harmonics \cite{jacobi}, \cite{addition} which will be employed later in the derivation of the SCF. When $\gamma$ is defined as,
\setcounter{equation}{19}
\begin{align}
\cos \gamma &=\cos\theta_{1}\cos\theta_{2}+\sin\theta_{1}\sin\theta_{2}\cos(\phi_{1}-\phi_{2}),
\end{align}
where $(\theta_{1},\phi_{1})$ and $(\theta_{2},\phi_{2})$ are the spherical coordinates of the vectors $\hat{\textbf{v}}$ and $\textbf{x}$ respectively, then the Legendre polynomial of argument cos($\gamma$) is given by (\ref{legend}), where  $P_{n}^{m}$ are the associated Legendre polynomials.

\subsection{Spatial Correlation Function Using SHE of Plane Waves}
These results are now employed to derive a closed-form expression for the correlation between Tx antenna ports. The SHE result for plane waves in (\ref{Jacobi}) yields alternate expressions for the array responses of Tx and Rx antenna ports. It can be seen from Fig. \ref{antenna} that for Tx antenna ports placed along $\hat{\textbf{e}}_{y}$ direction with $||\textbf{x}||$ = $d_{t}|s-s'|$, $\hat{\textbf{x}}.\hat{\textbf{v}}$ is given by $\sin\phi\sin\theta$. Therefore the array response of the $s^{th}$ Tx antenna port can be expressed alternatively as (\ref{exp}).

Also from Fig. \ref{antenna}, the spherical coordinates $(\theta_{1},\phi_{1})$ of the wave vector $k\hat{\textbf{v}}$ are $(\theta,\phi)$, and the spherical coordinates $(\theta_{2},\phi_{2})$ of $\textbf{x}$ i.e., the vector along  $d_{t}(s-s')$ are $(\frac{\pi}{2},\frac{\pi}{2})$. Combining the addition theorem in (\ref{legend}) with (\ref{exp}), such that $\cos\gamma$=$\sin\phi\sin\theta$ and using the resulting expression in (\ref{Tx}) would expand $\rho_{t}(s-s')$ to yield (\ref{int}).

\textit{Proposition 1:} For a uniform linear array of antenna ports with arbitrary antenna patterns and for arbitrary angular distributions such that the $\phi,\varphi \in [-\pi,\pi]$ and $\theta, \vartheta \in [0,\pi]$, the correlation between any pair of Tx antennas ports can be expanded in a systematic way to yield (\ref{prop}), where $\bar{P}_{n}^{m}$(x)=$\sqrt{(n+\frac{1}{2})\frac{(n-m)!}{(n+m)!}}P_{n}^{m}(\text{x})$ and $\beta_{t}=\frac{2\pi}{\lambda}d_{t}$.

The proof of Proposition 1 follows from the following properties of Legendre and associated Legendre polynomials, 
\\
1) $P_{n}(0)=0$ if $n$ is odd, \\
2) $P_{n}^{m}(0)=0$ if ($n+m$) is odd, \\
3) $P_{0}(x)=1$. \\
The following trigonometric relations were also used, 
\\
1) $\cos(m'(\phi-\frac{\pi}{2}$)) =$(-1)^{m}\cos(2m\phi)$ for $m'=2m$,  \\
2) $\cos(m'(\phi-\frac{\pi}{2}$)) =$(-1)^{m-1}\sin((2m-1)\phi)$ for $m'=2m-1$, \\
 where $m=1, \dots, n$.

Defining $\bar{P}_{n}^{m}$(x)=$\sqrt{(n+\frac{1}{2})\frac{(n-m)!}{(n+m)!}}P_{n}^{m}(\text{x})$ and using the properties just described, (\ref{int}) can be expanded in a systematic way. Finally, using the decomposition of $g_{t}$($\phi,\theta,\theta_{tilt}$) $\approx$ $g_{t,H}(\phi)g_{t,V}(\theta,\theta_{tilt})$, and taking the deterministic terms out of the expectation operators yields (\ref{prop}).

The same approach would yield a similar expression for $\rho_{r}(u-u')$ with $g_{t}(\phi,\theta,\theta_{tilt})$ replaced by $g_{r}(\varphi,\vartheta)$ and the AoDs replaced by AoAs. The expansion looks alarming but will be shown to yield an interesting closed-form expression.

\subsection{Closed-form Expression for SCF in terms of Fourier Series coefficients of PAS and PES}
The expansion in (\ref{prop}) exhibits several difficulties in deriving a closed-form expression for the SCF. The random variables, AoD and AoA, with respect to which the expectations need to be computed appear as the arguments of Legendre polynomials. Several tables of Legendre and associated Legendre polynomials exist that express the first few Legendre polynomials as functions of its arguments. However, we need a general representation that can be used for any order to facilitate the development of the expectation terms. For this purpose, we use the trigonometric expansion of Legendre polynomials presented in \cite{legendre}.  The following Lemma expresses the Legendre and re-normalized associated Legendre polynomials with even and odd orders as a linear combination of sines and cosines.

\begin{figure*}[!t]
\normalsize
\setcounter{equation}{26}
\begin{align}
\label{tcorr}
&\rho_{t}(s-s')=\pi^{2}a_{\phi}(0)b_{\theta}(1)j_{0}\left(\beta_{t}|s-s'|\right)  + \sum_{n=1}^{\infty} (-1)^{n}(4n+1)j_{2n}\left(\beta_{t}|s-s'|\right){P}_{2n}(0)a_{\phi}(0) \pi^{2} \sum_{k=-n}^{n}p_{n-k}p_{n+k} \frac{1}{2} [b_{\theta}(2k+1)  \nonumber \\  
&- b_{\theta}(2k-1)]  + \sum_{n=1}^{\infty}4 (-1)^{n} j_{2n}\left(\beta_{t}|s-s'|\right)  \left(\sum_{m=1}^{n} (-1)^{m}  \bar{P}_{2n}^{2m} (0) a_{\phi}(2m) \pi^{2}  \sum_{k=0}^{n}c_{2n,2k}^{2m}  \frac{1}{2}[b_{\theta}(2k+1)-b_{\theta}(2k-1)] \right)   \\
&+\sum_{n=1}^{\infty}4 i (-1)^{n} j_{2n-1}\left(\beta_{t}|s-s'|\right)  \Big(\sum_{m=1}^{n} (-1)^{m}  \bar{P}_{2n-1}^{2m-1} (0) b_{\phi}(2m-1) \pi^{2} \sum_{k=1}^{n}d_{2n-1,2k-1}^{2m-1}  \frac{1}{2}[a_{\theta}(2k-2)-a_{\theta}(2k)]  \Big)  \nonumber
\end{align}
\begin{align}
\label{rcorr}
&\rho_{r}(u-u')=\pi^{2}a_{\varphi}(0)b_{\vartheta}(1)j_{0}\left(\beta_{r}|u-u'|\right)  + \sum_{n=1}^{\infty} (-1)^{n}(4n+1)j_{2n}\left(\beta_{r}|u-u'|\right){P}_{2n}(0)a_{\varphi}(0) \pi^{2} \sum_{k=-n}^{n}p_{n-k}p_{n+k} \frac{1}{2} [b_{\vartheta}(2k+1)  \nonumber \\  
&- b_{\vartheta}(2k-1)]  + \sum_{n=1}^{\infty}4 (-1)^{n} j_{2n}\left(\beta_{r}|u-u'|\right)  \left(\sum_{m=1}^{n} (-1)^{m}  \bar{P}_{2n}^{2m} (0) a_{\varphi}(2m) \pi^{2} \sum_{k=0}^{n}c_{2n,2k}^{2m}  \frac{1}{2}[b_{\vartheta}(2k+1)-b_{\vartheta}(2k-1)] \right)   \\
&+\sum_{n=1}^{\infty}4 i (-1)^{n} j_{2n-1}\left(\beta_{r}|u-u'|\right)  \Big(\sum_{m=1}^{n} (-1)^{m}  \bar{P}_{2n-1}^{2m-1} (0) b_{\varphi}(2m-1) \pi^{2} \sum_{k=1}^{n}d_{2n-1,2k-1}^{2m-1}  \frac{1}{2}[a_{\vartheta}(2k-2)-a_{\vartheta}(2k)]  \Big)  \nonumber
\end{align}
\hrulefill
\vspace*{4pt}
\end{figure*}

\textit{Lemma 1} (From \cite{legendre}): For non-negative integers $n$ and $m$,
\setcounter{equation}{24}
\begin{align}
\label{legendre1}
P_{2n}(\cos x)&=p_{n}^{2}+2\sum_{k=1}^{n} p_{n-k} p_{n+k} \cos(2kx),  \nonumber \\
\bar{P}_{2n}^{2m}(\cos x)&=\sum_{k=0}^{n} c_{2n,2k}^{2m} \cos(2kx),   \\
\bar{P}_{2n-1}^{2m-1}(\cos x)&=\sum_{k=1}^{n} d_{2n-1,2k-1}^{2m-1} \sin((2k-1)x), \nonumber
\end{align}
where the coefficients $p_{n},c_{2n,2k}^{2m}$ and $d_{2n-1,2k-1}^{2m-1}$ are generated using recursion relations in [\cite{legendre}, equations 2.8, 3.1-3.5].  

This Lemma is the most important ingredient in the derivation of the SCF that gives it its generalized form that has not been derived before.   We now state Theorem 1 that describes how the correlation between any pair of channels constituted by distinct pairs of Tx and Rx antenna ports can be computed. 

\textit{Theorem 1:} For a uniform linear array of antenna ports with arbitrary antenna patterns and for arbitrary angular distributions such that the $\phi,\varphi \in [-\pi,\pi]$ and $\theta, \vartheta \in [0,\pi]$, the SCF can be computed as,
\begin{align}
\label{theorem}
\text{SCF} &= \rho_{t}(s-s') \rho_{r}(u-u'),
\end{align} 
where $\rho_{t}(s-s')$ and $\rho_{r}(u-u')$ are given by (\ref{tcorr}) and (\ref{rcorr}) respectively, given that $a_{\phi}(k), b_{\phi}(k), a_{\theta}(k)$ and $b_{\theta}(k)$ are the FS coefficients of PAS and PES respectively defined as,
\setcounter{equation}{28}
\begin{align}
a_{\phi}(m)&= \frac{1}{\pi}\int_{-\pi}^{\pi}\text{PAS}_{t}(\phi)\cos(m\phi) d\phi,  \\
b_{\phi}(m)&= \frac{1}{\pi}\int_{-\pi}^{\pi}\text{PAS}_{t}(\phi)\sin(m\phi) d\phi,  \\
a_{\theta}(k)&=\frac{1}{\pi}\int_{0}^{2\pi}\text{PES}_{t}(\theta,\theta_{tilt})\cos(k\theta) d\theta ,  \\
b_{\theta}(k)&= \frac{1}{\pi}\int_{0}^{2\pi}\text{PES}_{t}(\theta,\theta_{tilt})\sin(k\theta) d\theta.
\end{align}
The proof of Theorem 1 is postponed to Appendix A. Theorem 1 describes a novel method for obtaining the spatial correlation coefficients for 3D MIMO channels for arbitrary choices of antenna patterns and angular distributions which is often a difficult task. The proposed method is unique and is in contrast to most of the previous works that assume an underlying angular distribution and form of antenna patterns. This derivation can be generalized to other antenna topologies as well.

\textit{Remark 1:} The proposed SCF in Theorem 1 involves an infinite summation over $n$. However this infinite summation can be truncated to a small finite  number, $N_{0}$, of terms such that the truncation error has a bound that decreases exponentially with $N_{0}$. This has been proved through an extensive analysis in \cite{intrinsic} and \cite{error}. The authors proved the bound for the 3D multipath field in (\ref{Jacobi}). We extend the analysis to the correlation  expressions in (\ref{tcorr}) and (\ref{rcorr}).  
\begin{align}
&\rho_{t}(s-s')=\sum_{n=0}^{\infty} i^{n}(2n+1)j_{n}(k||\textbf{x}||) \mathbb{E}\big[g_{t}(\phi,\theta, \theta_{tilt})P_{n}\left(\hat{\textbf{x}}.\hat{\textbf{v}} \right)\big], \\
&\epsilon_{N_{0}}=\sum_{n > N_{0}} i^{n}(2n+1)j_{n}(k||\textbf{x}||) \mathbb{E}\left[g_{t}(\phi,\theta, \theta_{tilt})P_{n}\left(\hat{\textbf{x}}.\hat{\textbf{v}} \right)\right] \nonumber 
\end{align}
\begin{align}
&\leq \sum_{n > N_{0}} (2n+1)|j_{n}(k||\textbf{x}||)| \hspace{.05 in} |\mathbb{E}\left[g_{t}(\phi,\theta, \theta_{tilt})P_{n}\left(\sin\phi \sin \theta \right)\right]| \nonumber \\
\label{error_bound}
&\leq \sum_{n > N_{0}} (2n+1)|j_{n}(k||\textbf{x}||)| G_{P,max}
\end{align}
since  $\sup_{|x| \leq 1} |P_{n}(x)| \leq 1$ \cite{legendre_bound}, leading to $|\mathbb{E}\left[g_{t}(\phi,\theta, \theta_{tilt})P_{n}\left(\sin\phi \sin \theta \right)\right]| \leq G_{P,max}$. (\ref{error_bound}) is the same as equation (10b) in \cite{error} upto a scaling factor. The analysis in \cite{error} then uses the bound on spherical Bessel function and the Stirling bound on the Gamma function to show that for a finite $||\textbf{x}||$, the multipath field and hence $\rho_{t}(s-s')$ and $\rho_{r}(u-u')$ can be truncated to $|n| \leq N_{0}$, such that the truncation error is bounded by
\begin{equation}
{\epsilon}_{N_{0}} \leq G_{P,max} \eta \exp(-\delta),
\end{equation}
where $\delta = N_{0}-\lceil e k ||\textbf{x}||/2 \rceil$, $\delta \geq 0$ and $\eta \approx .678481234$.
For our analysis, to calculate the correlation between adjacent Tx or Rx antennas for an antenna spacing of $||\textbf{x}||=.5 \lambda$, given that $G_{P,max}=17 \rm{dBi}$,  $N_{0}$=14 would suffice to bound the error by approximately 0.5 \%.

\begin{figure}[!t]
\centering
\includegraphics[width=2.6 in]{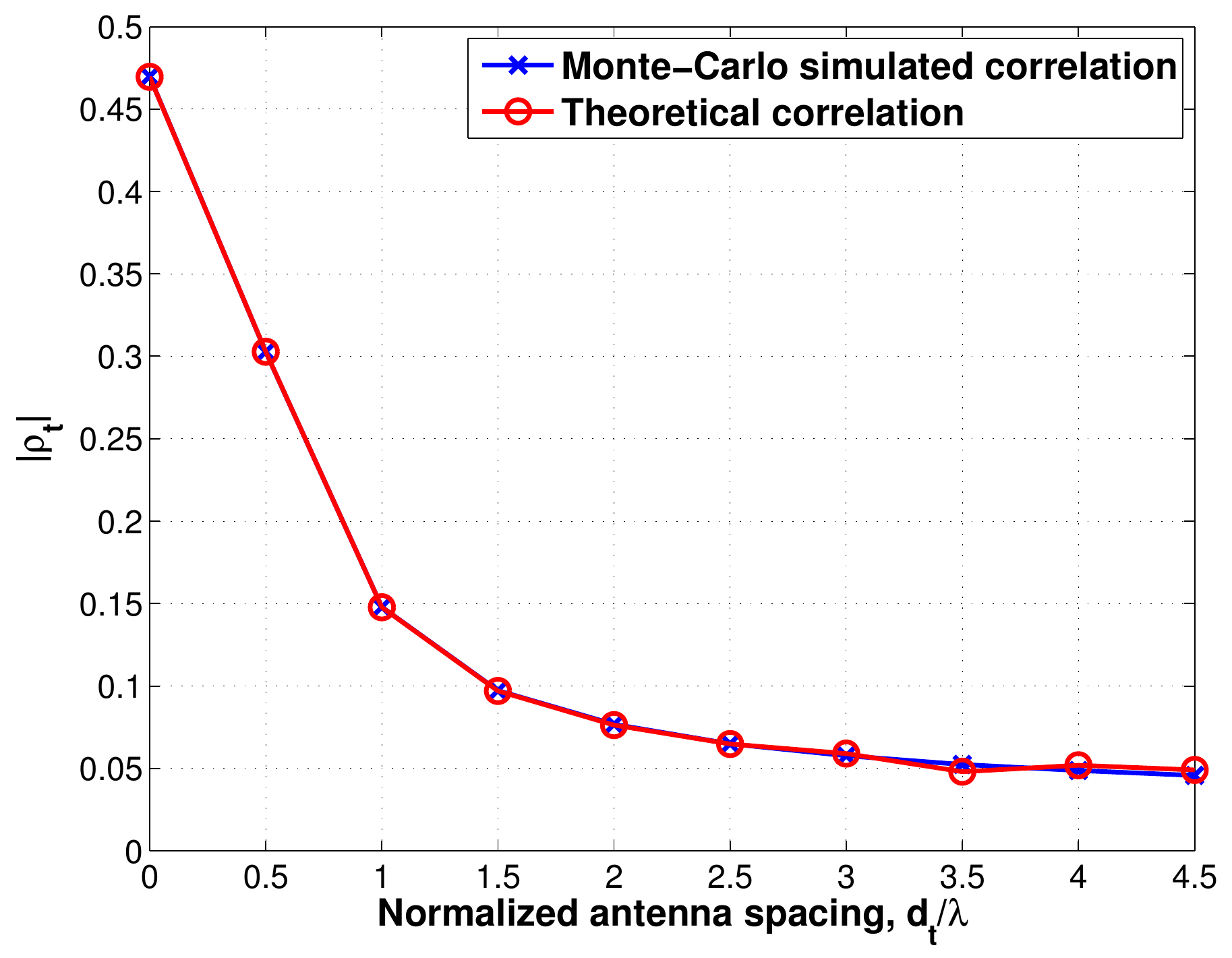}
\caption{Correlation between Tx antenna ports for distributions and patterns from standards.}
\label{Tstand}
\end{figure}
\begin{figure}[!t]
\centering
\includegraphics[width=2.6 in]{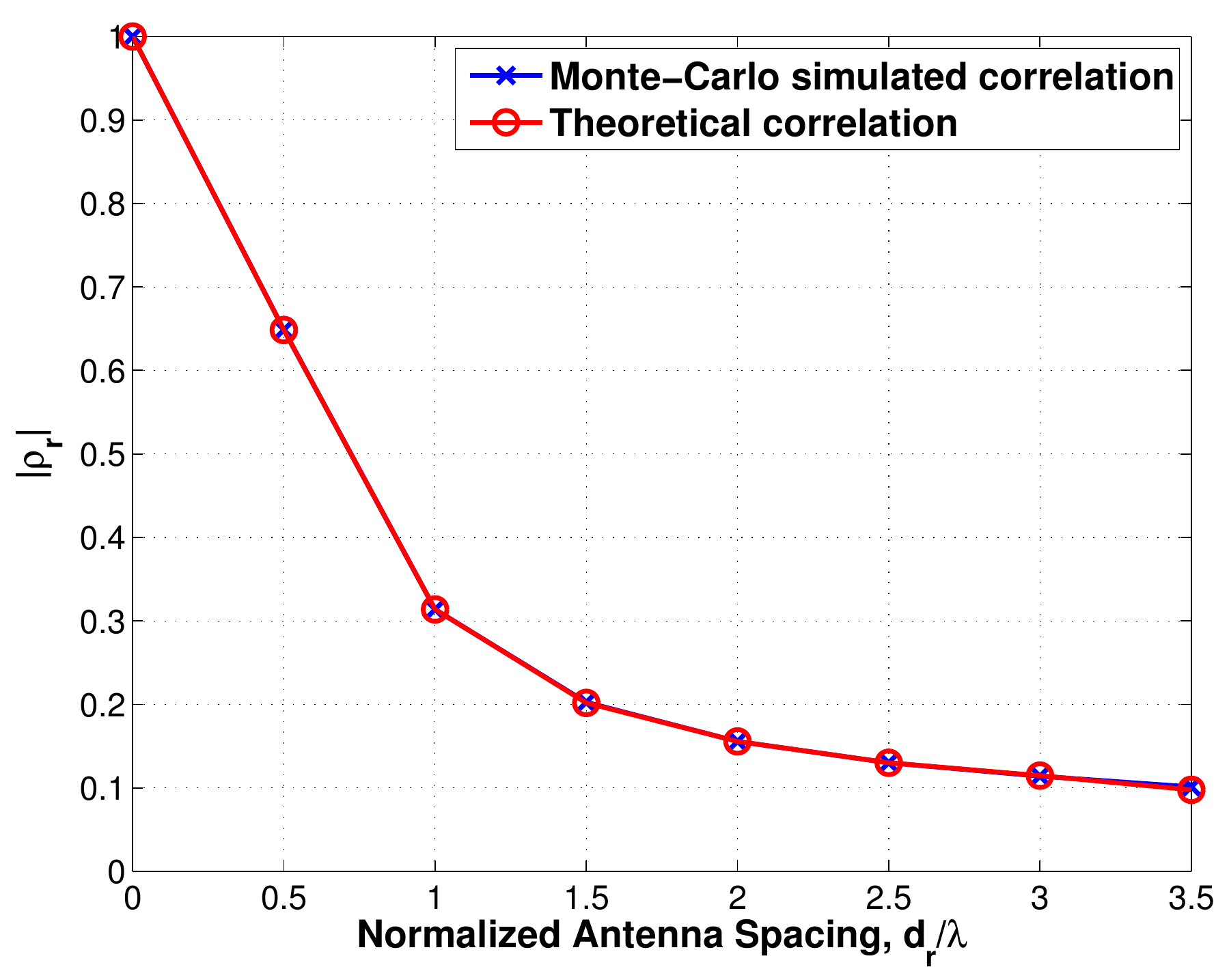}
\caption{Correlation between Rx antenna ports for distributions and patterns from standards.}
\label{Rstand}
\end{figure}


\section{Validation of the SCF for Standardized 3D Channel Model}
In the last section, we presented a generalized methodology for obtaining a closed-form expression for the spatial correlation function for the 3D channel, that works for any arbitrary choice of antenna patterns and distributions of azimuth and elevation AoD and AoA. Since this paper largely focuses on the guidelines provided in the mobile communication standards used globally, it is important that our model is validated using the angular distributions and antenna patterns specified in the standards. 

\begin{figure}[!t]
\centering
\includegraphics[width=2.6 in]{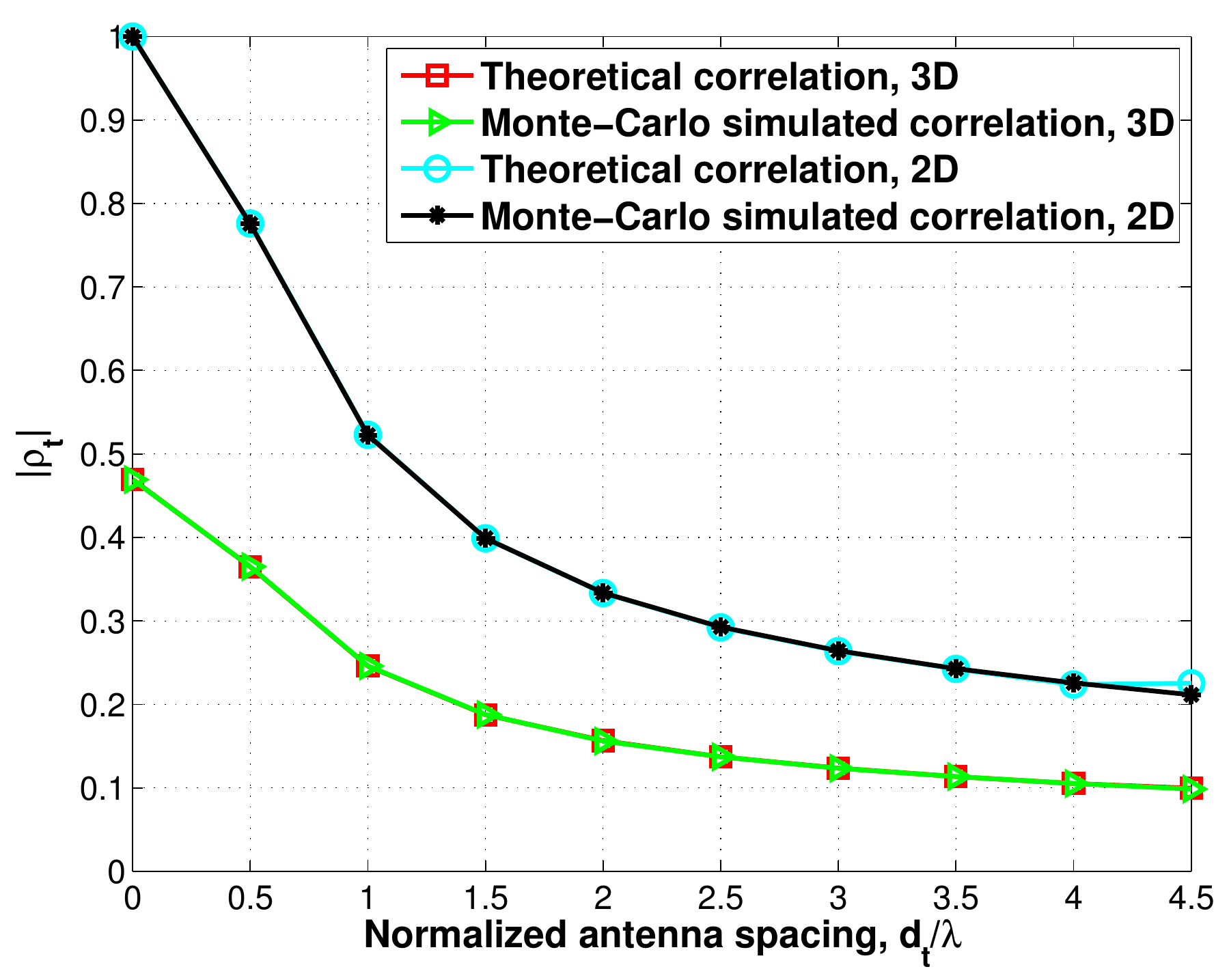}
\caption{Comparison of the correlation between Tx antenna ports for 2D and 3D channels.}
\label{without_ele}
\end{figure}

 In the standards, elevation AoD and AoA are drawn from Laplacian elevation density spectrum given in (\ref{laplace}), where $\theta_{0}$ is the mean AoD/AoA and $\sigma$ is the angular spread in the elevation.  The characteristics of azimuth angles are well captured by Wrapped Gaussian (WG) density spectrum \cite{ITU}, \cite{Kalliola02angularpower}, \cite{PES1}. However in the recent years, the Von Mises (VM) distribution has received great attention in modeling nonisotropic propagation due to its close association with the WG spectrum \cite{correlation2}, \cite{vonmises1}. This distribution given by,
\setcounter{equation}{35}
\begin{align}
\label{VMdist}
f_{\phi}(\phi)&=\frac{\exp(\kappa \cos(x-\mu))}{2\pi I_{0}(\kappa)}
\end{align}
is related to the WG distribution through a straightforward relationship obtained using their first circulant moments \cite{vonmises},
\begin{align}
& \text{WG}(\mu,\sigma^{2}) = \text{VM}(\mu,\kappa), \hspace{.01 in} \sigma^{2}=2[\log I_{0}(\kappa)-\log I_{1}(\kappa)],
\end{align}
where $I_{n}(\kappa)$ is the modified Bessel function of order $n$, $\mu$ is the mean AoD/AoA and $\frac{1}{\kappa}$ is a measure of dispersion. 

In practice, directional antenna patterns are commonly used so the incorporation of these patterns in this work is noteworthy. As per the standards, the global antenna port pattern is given by (\ref{port}) for each Tx port. However, in order to obtain a closed-form expression for the FS coefficients of PAS using VM distribution, the antenna ports are assumed to be omnidirectional in the azimuth, i.e. $g_{t,H}(\phi)=0$ $\rm{dB}$. From a propagation viewpoint, it is the downtilt angle that determines the vertical pattern which would play a crucial role in determining the antenna directivity and its potential to change our perception on the physical distribution of scatterers. Therefore assuming non-isotropy in the azimuth will not affect the results to a considerable extent. The vertical antenna pattern $g_{t,V}(\theta,\theta_{tilt})$ is given by (\ref{vpattern}). The pattern at the MS, $g_{r}(\varphi, \vartheta)=0$ $\rm{dB}$ in the standards because mobile terminals should not, in most cases, favor any direction. Using these antenna patterns and angular densities, the FS coefficients of PAS and PES are computed and provided in Appendix B. 

\begin{figure}[!t]
\centering
\includegraphics[width=2.6 in]{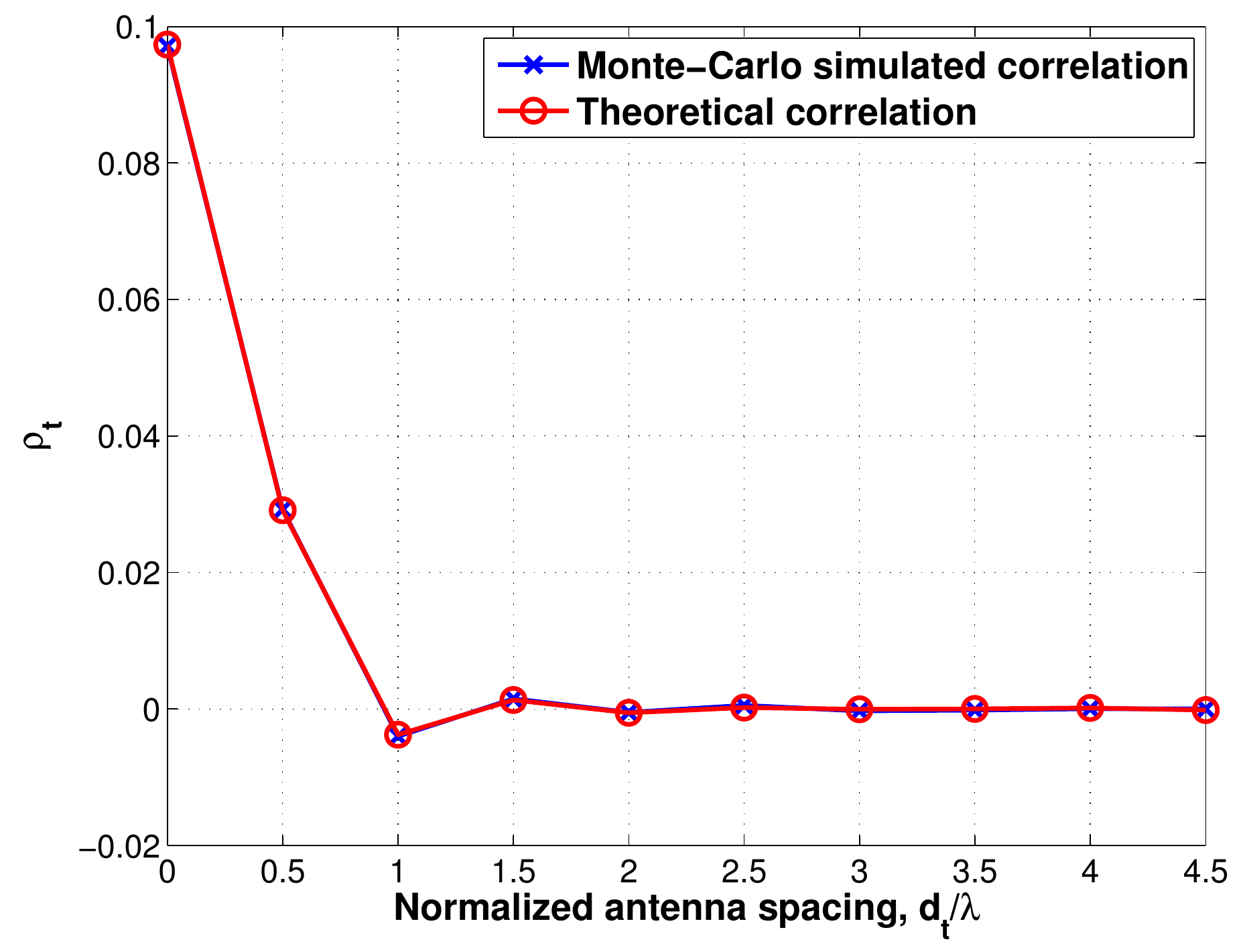}
\caption{Correlation between Tx antenna ports for uniform azimuth angular distribution and patterns from standards.}
\label{uniform}
\end{figure}

The validation of the theoretical results in (\ref{tcorr}) and (\ref{rcorr}) is done by comparison with Monte-Carlo simulation results. The Monte-Carlo simulations are performed over two thousand channel realizations using (\ref{channel}) to compute the correlation values in (\ref{Tx}) and (\ref{Rx}). The angles of the propagation paths are generated using the densities described earlier and the antenna patterns from the standards are used for the purpose of validation. The Monte-Carlo simulated correlation between Tx antennas and Rx antennas is then compared with the theoretical correlation computed using (\ref{tcorr}) and (\ref{rcorr}) respectively. The theoretical results should provide a very accurate fit to the Monte-Carlo simulated correlation. For the simulations, we set $N_{0}=15$, $\theta_{tilt}=95^{o}$, $\theta_{3dB}=15^{o}$, $\phi_{3dB}=70^{o}$, $\sigma_{s}=7^{o}$, $\sigma_{u}=10^{o}$, $\theta_{0}=90^{o}$, $\kappa_{s}, \kappa_{u}=5$ and $\mu=\frac{2\pi}{3}$. The results are shown in Fig. \ref{Tstand} and Fig. \ref{Rstand}. As expected, the correlation is seen to decrease as the distance between the pair of antenna ports increases and more importantly, it is apparent from the graphs that the theoretical results provide a perfect fit to the Monte-Carlo simulated correlation for as few as fifteen summations over $n$. 

A comparison is made against the 2D standardized model given by, 
\begin{align}
\label{2Dchannel}
[\textbf{H}]_{su}&= \sum\limits_{n=1}^N  \alpha_{n} \sqrt{g_{t,H}(\phi_{n})} \exp \left(i k (s-1) d_{t}\sin\phi_{n}\right)  \nonumber \\ 
& \sqrt{g_{r,H}(\varphi_{n})} \exp\left(i k  (u-1) d_{r}\sin\varphi_{n}\right), 
\end{align}
Note that this is obtained by using $\theta$, $\vartheta=\pi/2$ and $g_{t,V}(\theta,\theta_{tilt})$, $g_{r,V}(\vartheta)=0$ $\rm{dB}$ in (\ref{channel}). The correlation between the Tx antennas is then given by, 
\begin{align}
\label{tcorr_2D}
\rho_{t}(s-s')&=\mathbb{E}\left[g_{t,H}(\phi)\exp\left(i\frac{2\pi}{\lambda}d_{t}(s-s')\sin\phi\right)\right]. 
\end{align}
The same theoretical analysis done for the 3D channel can be repeated here and would lead to the squaring of $P_{2n}(0)$, $\bar{P}_{2n}^{2m}(0)$ and $\bar{P}_{2n-1}^{2m-1}(0)$ and the removal of all expectation terms involving the elevation angles in (\ref{prop}). The theoretical result for (\ref{tcorr_2D}) is then obtained straightforwardly. The Monte-Carlo simulated correlation in (\ref{tcorr_2D}) and the theoretical result are plotted in Fig. \ref{without_ele}. The 2D model clearly overestimates the correlation, which is a consequence of ignoring the directivity of antennas and the propagation of multipaths in the elevation. Assuming the radiation of energy from all the antennas to be in the same fixed direction in the elevation will cause the antennas to appear more correlated.  

The developed results are further validated by using uniform angular distribution for azimuth angles. Both the horizontal and vertical antenna patterns, i.e. $g_{t,H}(\phi)$ and $g_{t,V}(\theta,\theta_{tilt})$ in (\ref{hpattern}) and (\ref{vpattern}) respectively, are considered this time. The correlation obtained is real which results from the symmetry of the uniform distribution and horizontal antenna pattern causing $b_{\phi}(k)$'s to be equal to zero. Fig. \ref{uniform} illustrates the excellent agreement between our derived and Monte-Carlo simulated results and establishes the credibility of the proposed function. 


\section{Equivalent Channel Model and Mutual Information Analysis}

\begin{table}
\normalsize
\centering
\caption{List of symbols in section V.}
\begin{tabular}{|c|c|}
\hline
Symbol & Description \\ 
\hline
 $\textbf{R}_{BS}$ & Correlation matrix at BS.  \\ 
$\textbf{R}_{MS}$ & Correlation matrix at MS.   \\ 
 $\textbf{X}$ & $N_{MS} \times N_{BS}$ matrix of i.i.d $\mathcal{CN} (0,1)$ entries. \\ 
 $\textbf{y}$, $\textbf{x}$ & Rx and Tx signals respectively.  \\ 
 $\textbf{n}$ & Received noise (AWGN).  \\ 
 $\sigma^{2}$ & Variance of \textbf{n}.  \\ 
$K$ & Number of users.  \\ 
$\textbf{G}$& RZF precoding matrix.  \\ 
 $\beta$ & Scaling  parameter of the precoder. \\ 
$\zeta$ & Regularization parameter of the precoder.  \\ 
 $\textbf{g}_{k}$ & Precoding  vector for the $k^{th}$ user. \\  
 $\textbf{h}_{k}$ & Channel vector for the $k^{th}$ user. \\  
  $s_{k}$ & Data symbol for the $k^{th}$ user. \\ 
  $\gamma_{k}$ & SINR for the $k^{th}$ user. \\ 
	$\gamma_{k}^{o}$ & Deterministic equivalent of SINR. \\ 
$\theta_{LoS,k}$ & Elevation line of sight angle at the BS.  \\
\hline
\end{tabular}
\label{Table2}
\end{table}

In this section, we focus on the MI analysis of the 3D channel model in order to quantify the effect of fading correlation and measure the performance gains realizable through potential elevation beamforming at the transmitter by careful selection of the downtilt angles. The important symbols used in this section are listed in Table \ref{Table2}.

\subsection{Kronecker Channel Model}
\begin{figure*}[!t]
\normalsize
\setcounter{equation}{43}
\begin{equation}
\label{V}
\begin{aligned}
V(\sigma^{2})= \frac{1}{N_{BS}}\log \det \left(\textbf{I} + \frac{1}{\sigma^{2}}\kappa(\sigma^{2}) \textbf{R}_{BS} \right)+\frac{1}{N_{BS}}\log \det \left(\textbf{I} + \frac{1}{\sigma^{2}}\bar{\kappa}(\sigma^{2}) \textbf{R}_{MS} \right)-\frac{1}{\sigma^{2}}\kappa(\sigma^{2}) \bar{\kappa}(\sigma^{2}),
\end{aligned}
\end{equation}
\hrulefill
\end{figure*}

The idea that multiple antennas at transmitter and receiver can bring about remarkable improvements in the MI made MIMO methods exceedingly popular. However this improvement depends on the multipath richness since a large capacity gain can be realized in the presence of potentially decorrelated channel coefficients. It is therefore, important for channel models to take this correlation into account to allow for a more accurate performance analysis. There are two popular approaches to model these correlated MIMO channels. The first one is the parametric approach, which was discussed in detail in Section II-A. This approach takes into account the spatial characteristics of wireless channels quite meticulously. The second approach is nonparametric, wherein the spatial correlation in the MIMO channel is reproduced using theoretical Tx and Rx spatial correlation matrices. The latter is more suitable for the information-theoretic analysis of MI. One of the widely used nonparametric channel models is the Kronecker model that relies on the two matrices describing the correlation characteristics at both ends of the communication link \cite{kronecker}, \cite{kronecker_model}, \cite{kronecker_model2}. This model is defined as,
\setcounter{equation}{39}
\begin{equation}
\label{Kronecker}
\begin{aligned}
\textbf{H} = \textbf{R}_{MS}^{\frac{1}{2}} \textbf{X} \textbf{R}_{BS}^{\frac{1}{2}} ,
\end{aligned}
\end{equation}
where $\textbf{X}$ is a $N_{MS}$x$N_{BS}$ matrix whose entries are independently and identically distributed according to a complex circularly symmetric Gaussian distribution, i.e. $\mathcal{C N}$(0,1), $\textbf{R}_{MS}$ is the correlation matrix at the MS with $[\textbf{R}_{MS}]_{u,u'}=\rho_{r}(u-u')$, $\textbf{R}_{BS}$ is the correlation matrix at the BS with $[\textbf{R}_{BS}]_{s,s'}=\rho_{t}(s-s')$ and $\rho_{t}(s-s')$, $\rho_{r}(u-u')$ are obtained using the derived expressions in (\ref{tcorr}) and (\ref{rcorr}). For antennas arranged in a linear array, $\textbf{R}_{BS}$ and $\textbf{R}_{MS}$ are Toeplitz. 

We consider the downlink of a single cell, where the BS is equipped with $N_{BS}$ antenna ports and the MS is equipped with $N_{MS}$ antenna ports. The channel is linear and time-invariant. A time-division duplex (TDD) protocol is considered where BS acquires instantaneous CSI in the uplink and uses it for downlink transmission by exploiting the channel reciprocity. The channel \textbf{H} is known only to the receiver but not to the transmitter. Therefore power is distributed equally over all Tx antennas instead of employing the water-filling scheme. Moreover \textbf{H} is fixed during the communication interval, so we do not need to time average the MI. The received complex baseband signal \textbf{y} $\in \mathbb{C}^{N_{MS} \times 1}$ at the MS is given by,
\begin{equation}
\textbf{y}=\textbf{H}\textbf{x}+\textbf{n},
\end{equation}
where \textbf{x} $\in \mathbb{C}^{N_{BS} \times 1}$ is the Tx signal from the BS, \textbf{H} is the $N_{MS}$x$N_{BS}$ channel matrix generated using (\ref{Kronecker}) and $\textbf{n}^{N_{MS}\times 1}$ is the additive white Gaussian noise (AWGN) with variance $\sigma^{2}$. The MI of the $N_{MS}$x$N_{BS}$ MIMO system with equal power-allocation is then given by,
\begin{equation}
\begin{aligned}
\label{MI}
I(\sigma^2)=\text{log} \hspace{.005 in}\ \text{det} \left(\textbf{I}_{N_{MS}}+ \frac{1}{N_{BS} \hspace{.03 in} \sigma^{2}} \textbf{H} \textbf{H}^{H} \right),
\end{aligned}
\end{equation}
where $\sigma^{2}$ is the noise variance and the average total Tx power is assumed to be 1.

The subsequent analysis in the next two subsections will make use of the deterministic equivalents of the MI available in literature to study the behavior of the correlated 3D MIMO channels through numerical results. In the numerical results that follow hereafter, we set $N_{BS}=20$, $N_{MS}=20$, $N_{0}=15$, $\theta_{tilt}=95^{o}$, $\theta_{3dB}=15^{o}$, $\phi_{3dB}=70^{o}$, $\sigma_{s}=3^{o}$, $\sigma_{u}=10^{o}$, $\theta_{0}=90^{o}$, $\kappa_{s}=5$, $\kappa_{u}=5$ and $\mu=0$. For the mono-user systems, we work for the SNR level of 0 $\rm{dB}$. Two thousand independent Monte-Carlo realizations of the parametric channel from (\ref{channel}) are generated to compute the MI values in (\ref{MI}) to allow for comparison with the theoretical results that follow. For these simulations, the azimuth angles are again drawn from VM distribution given in (\ref{VMdist}) and the elevation angles are drawn from Laplacian distribution given in (\ref{laplace}). The vertical antenna pattern given in  (\ref{vpattern}) is considered but again the antennas are assumed to be omnidirectional in the azimuth to allow for the calculation of FS coefficients of PAS in a closed-form. The theoretical correlation coefficients are calculated using (\ref{tcorr}) and (\ref{rcorr}).  The developed SCF is used to determine the covariance matrices at the transmitter and receiver which are needed for the Kronecker model in (\ref{Kronecker}).

\subsection{Mutual Information Analysis of a Mono-User System}
It is imperative to study the behaviour of MI of MIMO channels in the presence of fading correlation to evaluate different beamforming techniques. The MI for every realization of the channel can be viewed as a RV and it is interesting to study the statistics and distribution of this RV. Deriving closed-form expressions for the distribution of MI of the Kronecker model is a challenging task. However in the large ($N_{BS},N_{MS}$) regime, RMT provides some simple deterministic approximations to this distribution. These deterministic equivalents are quite accurate even for a moderate number of antennas. The deterministic equivalent for the MI of Kronecker model was studied by \cite{kronecker_model}, where it was shown that $\frac{I(\sigma^2)}{N_{BS}}$ converges to a deterministic quantity defined as the fixed point of an integral equation. More recently, Hachem \textit{et al} derived a deterministic equivalent for the MI of Kronecker channel model and rigorously proved that the MI converges to a standard Gaussian random variable in the asymptotic limit in \cite{walid_mutual_information}. The result is stated here and will be used in our analysis.

\begin{figure}[!t]
\centering
\includegraphics[width=2.65 in]{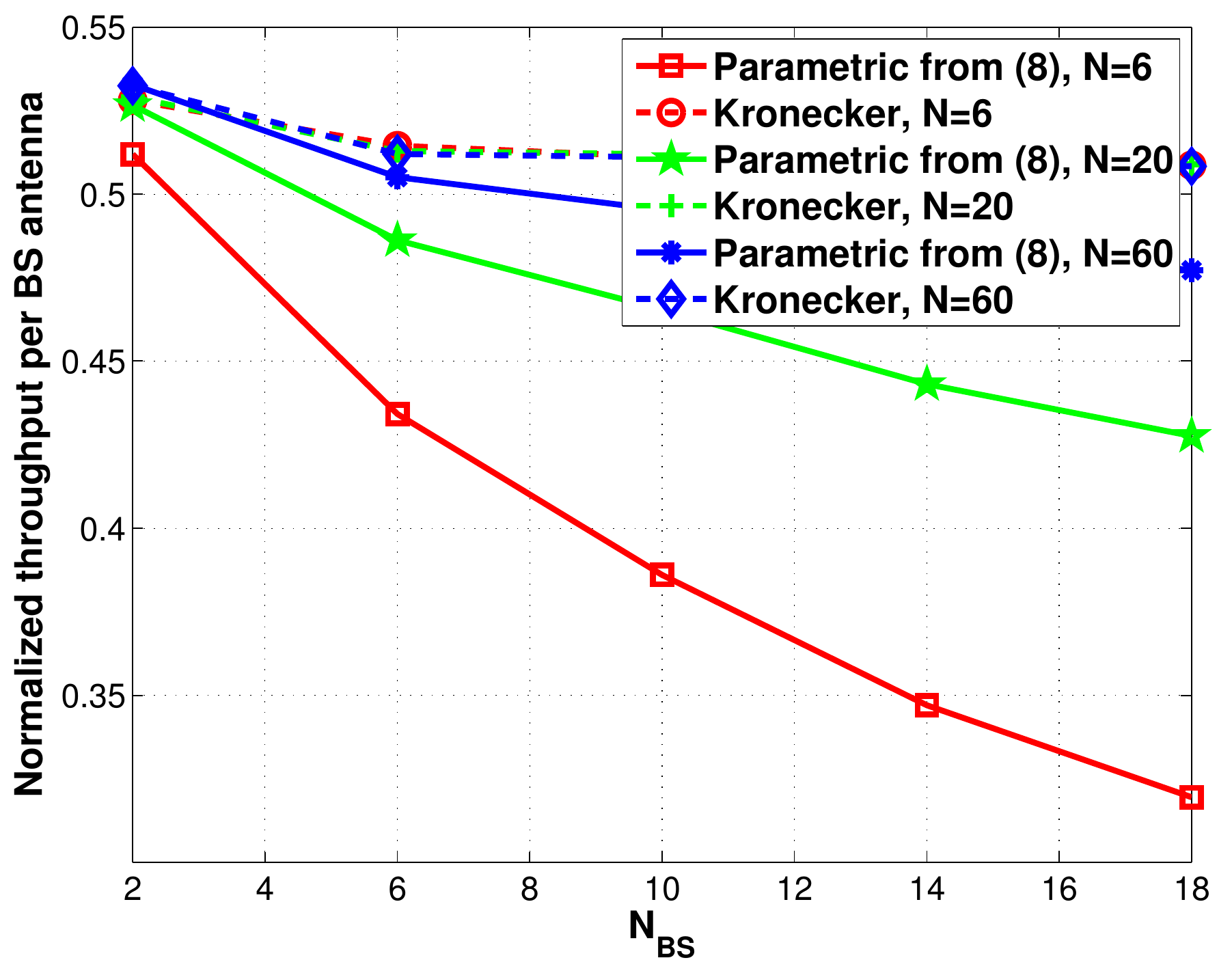}
\caption{Pinhole effect.}
\label{pinhole}
\end{figure}

Theorem 1 in \cite{walid_mutual_information}, Theorem 4.1. in \cite{walid_mutual_information1} suggests that for the Kronecker channel model given by \textbf{H}=$\textbf{R}_{MS}^{\frac{1}{2}} \textbf{X} \textbf{R}_{BS}^{\frac{1}{2}}$, the ergodic MI converges to a deterministic quantity, given that the assumptions \textbf{A1} and \textbf{A2} in \cite{walid_mutual_information} on the eigenvalues of $\textbf{R}_{BS}$ and $\textbf{R}_{MS}$ hold. Under this setting, the authors showed,
\begin{equation}
\begin{aligned}
\label{walideq}
\frac{1}{N_{BS}} \mathbb{E} [I(\sigma^{2})]-V(\sigma^{2}) \xrightarrow[N_{BS},N_{MS}\rightarrow \infty]{a.s} 0,
\end{aligned}
\end{equation}
where $V(\sigma^{2})$ is given by (\ref{V}) and where, ($\kappa(\sigma^{2}),\bar{\kappa}(\sigma^{2})$) is the unique positive solution of the system of equations given by,
\setcounter{equation}{44}
\begin{align}
&\kappa(\sigma^{2})=\frac{1}{N_{BS}} \text{tr} \left( \textbf{R}_{MS}\left(\textbf{I}+\frac{1}{\sigma^{2}}\bar{\kappa}(\sigma^{2}) \textbf{R}_{MS} \right)^{-1} \right), \\
& \bar{\kappa}(\sigma^{2})=\frac{1}{N_{BS}} \text{tr} \left( \textbf{R}_{BS}\left(\textbf{I}+\frac{1}{\sigma^{2}}\kappa(\sigma^{2}) \textbf{R}_{BS} \right)^{-1} \right).
\end{align}

\begin{figure}[!t]
\centering
\includegraphics[width=2.65 in]{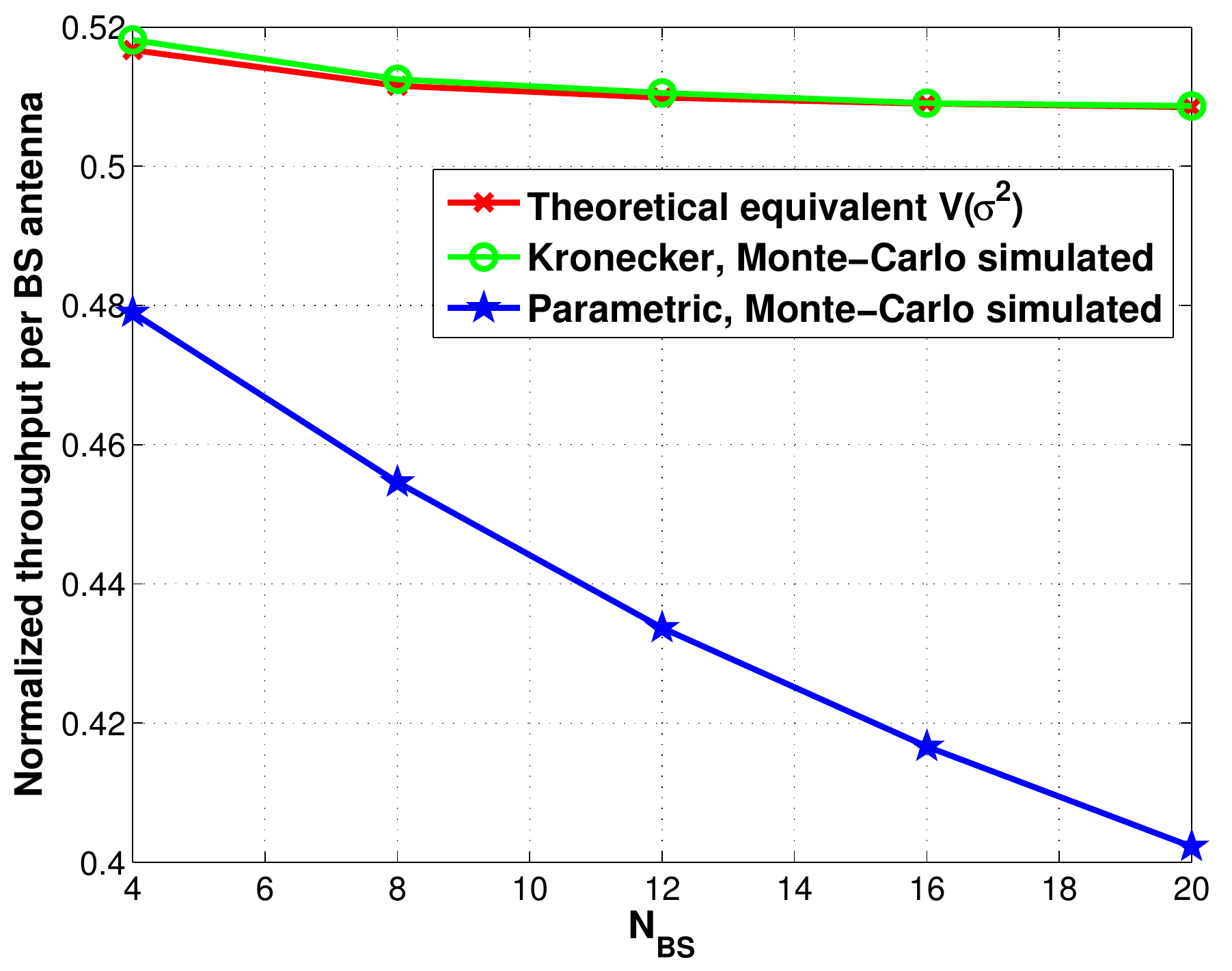}
\caption{Verification of deterministic equivalent of MI of Kronecker model.}
\label{detmono}
\end{figure}

\begin{figure}[!t]
\centering
\includegraphics[width=2.75 in]{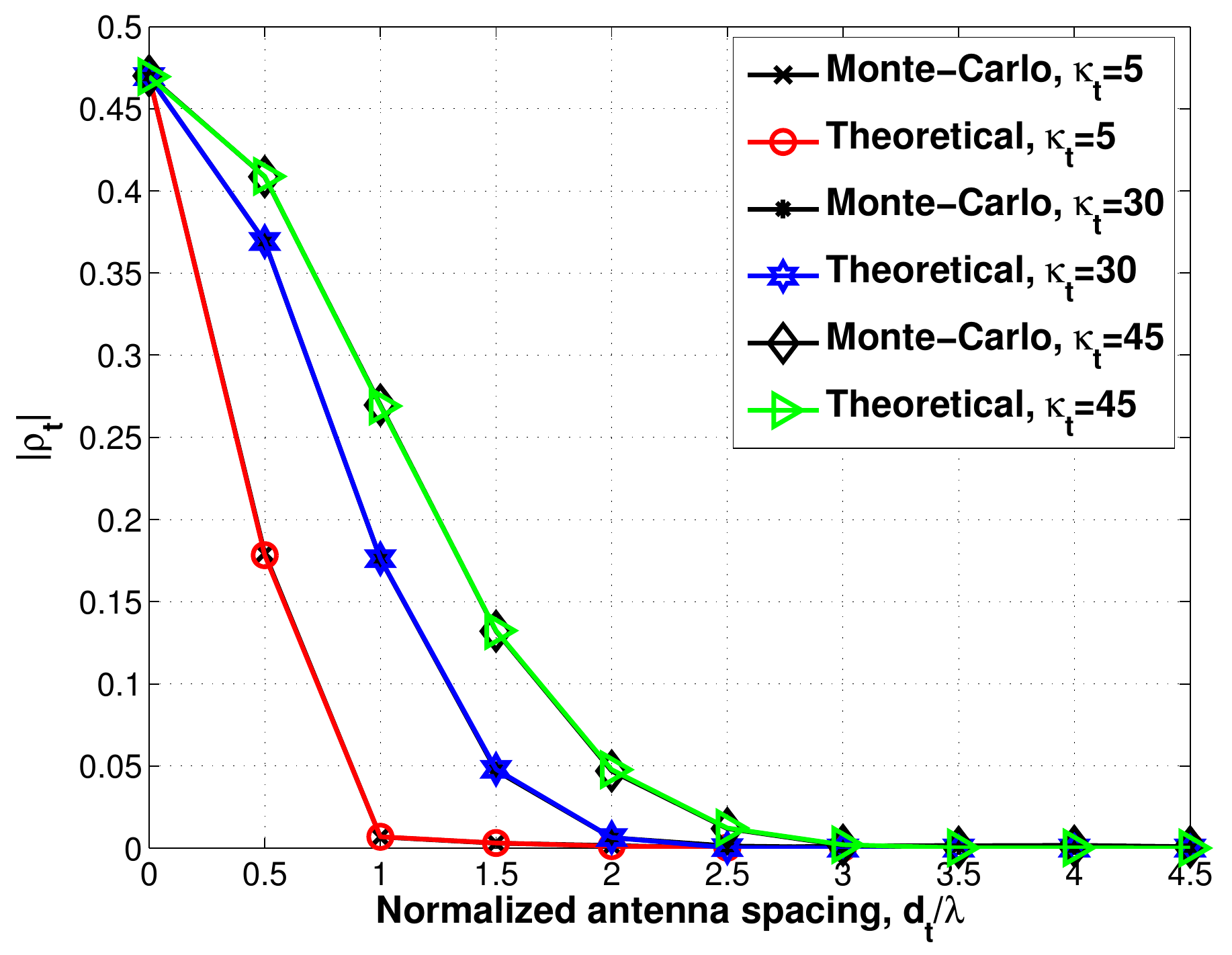}
\caption{Effect of azimuth angular spread on correlation.}
\label{kappa}
\end{figure}
\begin{figure}[!t]
\centering
\includegraphics[width=2.75 in]{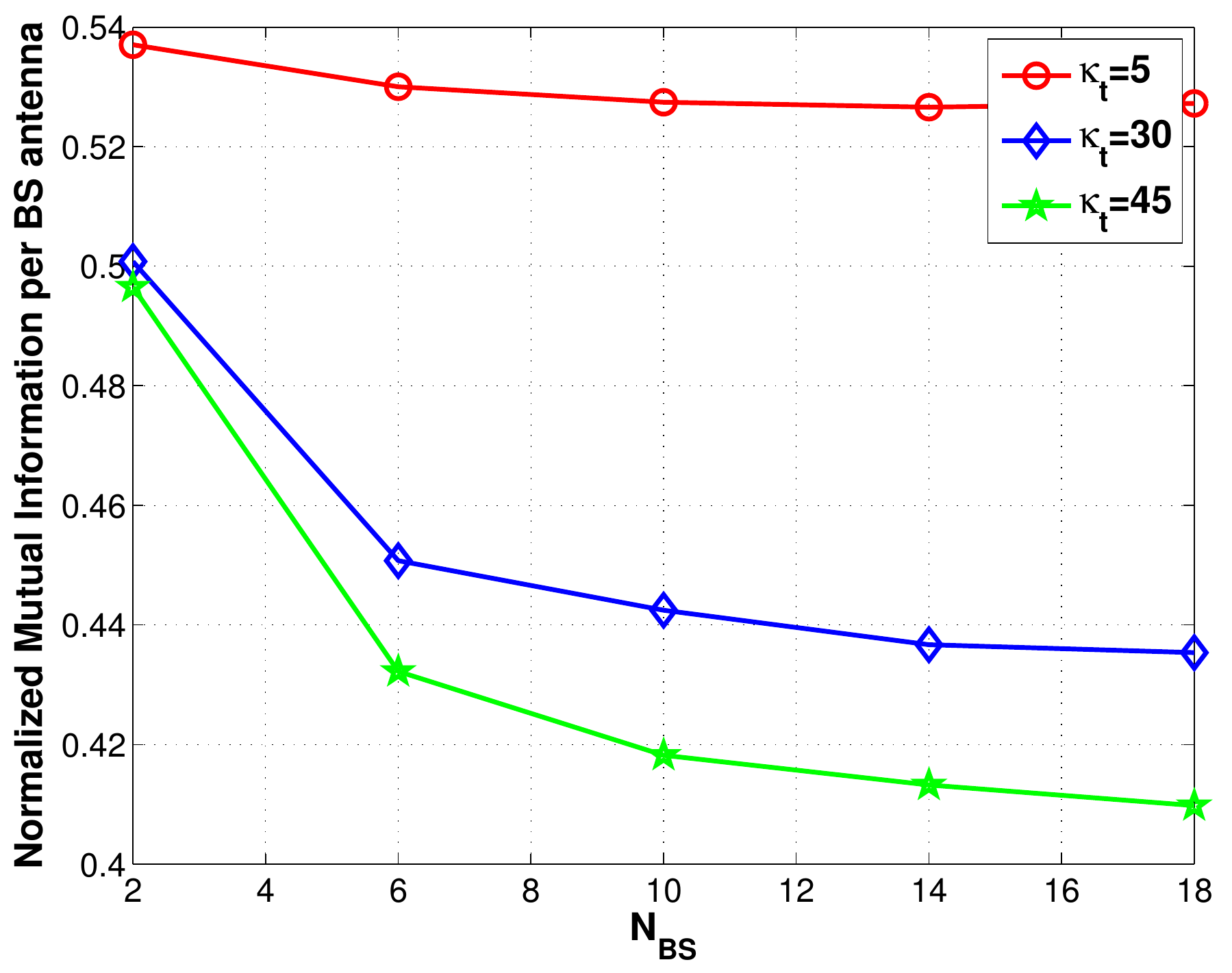}
\caption{Effect of azimuth angular spread on mutual information.}
\label{kappaMI}
\end{figure}

This theorem shows how the MI can be approximated by a deterministic equivalent in the asymptotic regime. The result remains relevant even for a moderate number of antennas as will be confirmed through simulations. Before discussing the results, it is crucial to point out a restriction to Kronecker channel model regarding the loss of information about the number of paths. The rank structure of MIMO channel matrix not only depends on the correlation present in the channel but also exhibits a strong dependence on the structure of scattering in the propagation environment. It is possible to have a rank deficient channel matrix even if the fading is decorrelated at both ends due to a small number of multipaths as compared to the number of antennas. This phenomenon, known as pinhole or keyhole effect, is generally observed in mild scattering conditions or when the communication link is very long. It has been studied in \cite{gesbert}, \cite{keyhole}. The pinhole effect is captured by the parametric channel models, like the one in (\ref{channel}), that explicitly depend on the number of propagation paths. However the nonparametric model in (\ref{Kronecker}) does not exhibit this phenomenon as there is no information about the number of multipaths involved. This has been illustrated in Fig. \ref{pinhole}. Monte-Carlo simulations are performed over two thousand channel realizations using (\ref{channel})  and (\ref{Kronecker}) to compute the MI values for the parametric channel model and non-parametric Kronecker channel model respectively  for different number of propagation paths $N$. It can be seen that a reduction in the number of paths can severely affect the channel MI as confirmed by the results for the parametric model. However, the MI obtained using the Monte-Carlo realizations of the Kronecker channel model is unaffected and only depends on the degree of spatial correlation present in the channel. Next we verify the deterministic equivalent for the MI of Kronecker channel model in (\ref{walideq}). The theoretical result in (\ref{V}) is plotted  in Fig. \ref{detmono} for $N=20$ and is seen to coincide quite perfectly with the MI obtained using the  Monte-Carlo realizations of the non-parametric Kronecker model in (\ref{Kronecker}), for a moderate number of antennas. The MI using the Monte-Carlo realizations of the parametric model in (\ref{channel}) is also plotted and the pinhole effect is again apparent. 

\begin{figure}[!t]
\centering
\includegraphics[width=2.75 in]{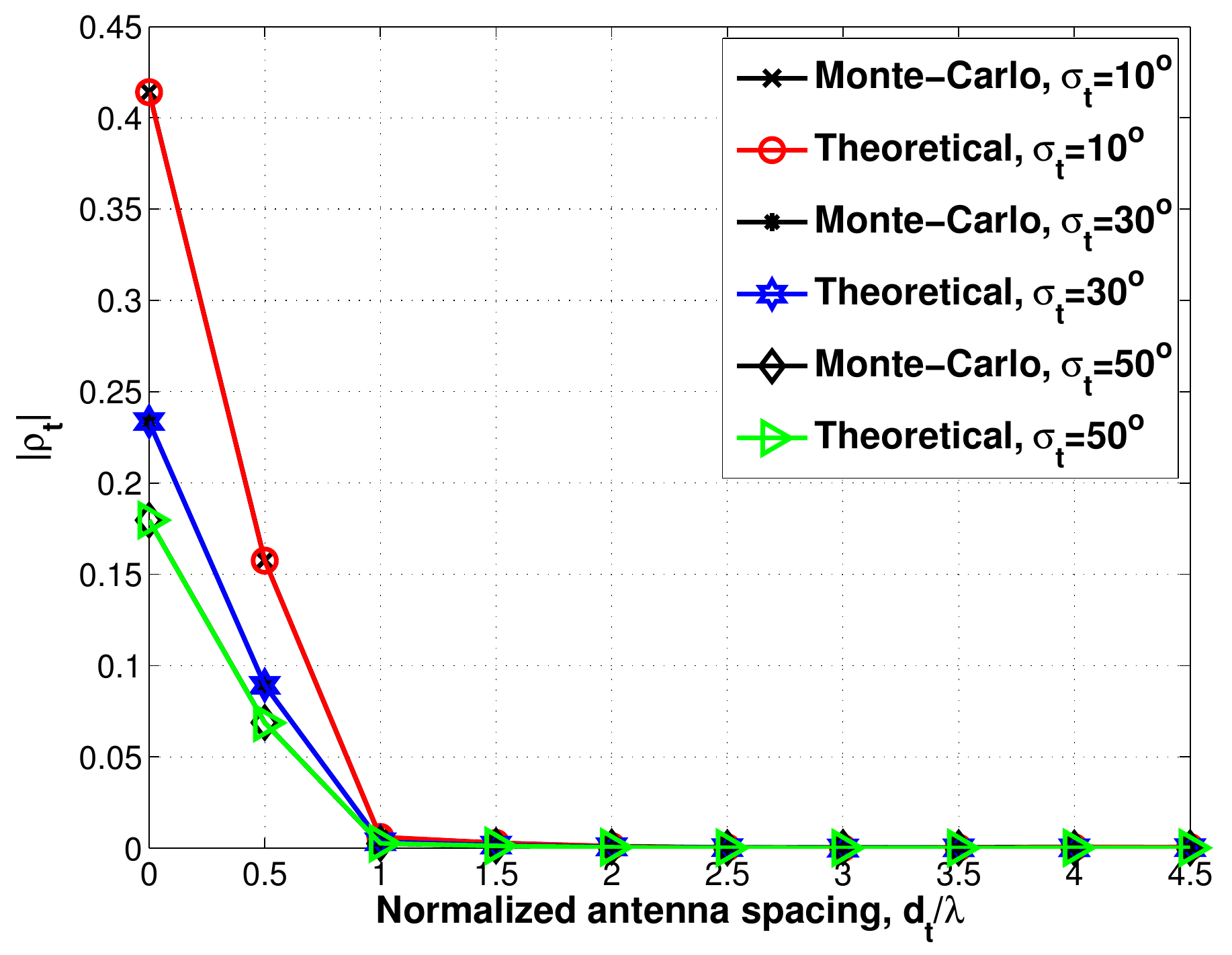}
\caption{Effect of elevation angular spread on correlation.}
\label{sigma}
\end{figure}
\begin{figure}[!t]
\centering
\includegraphics[width=2.75 in]{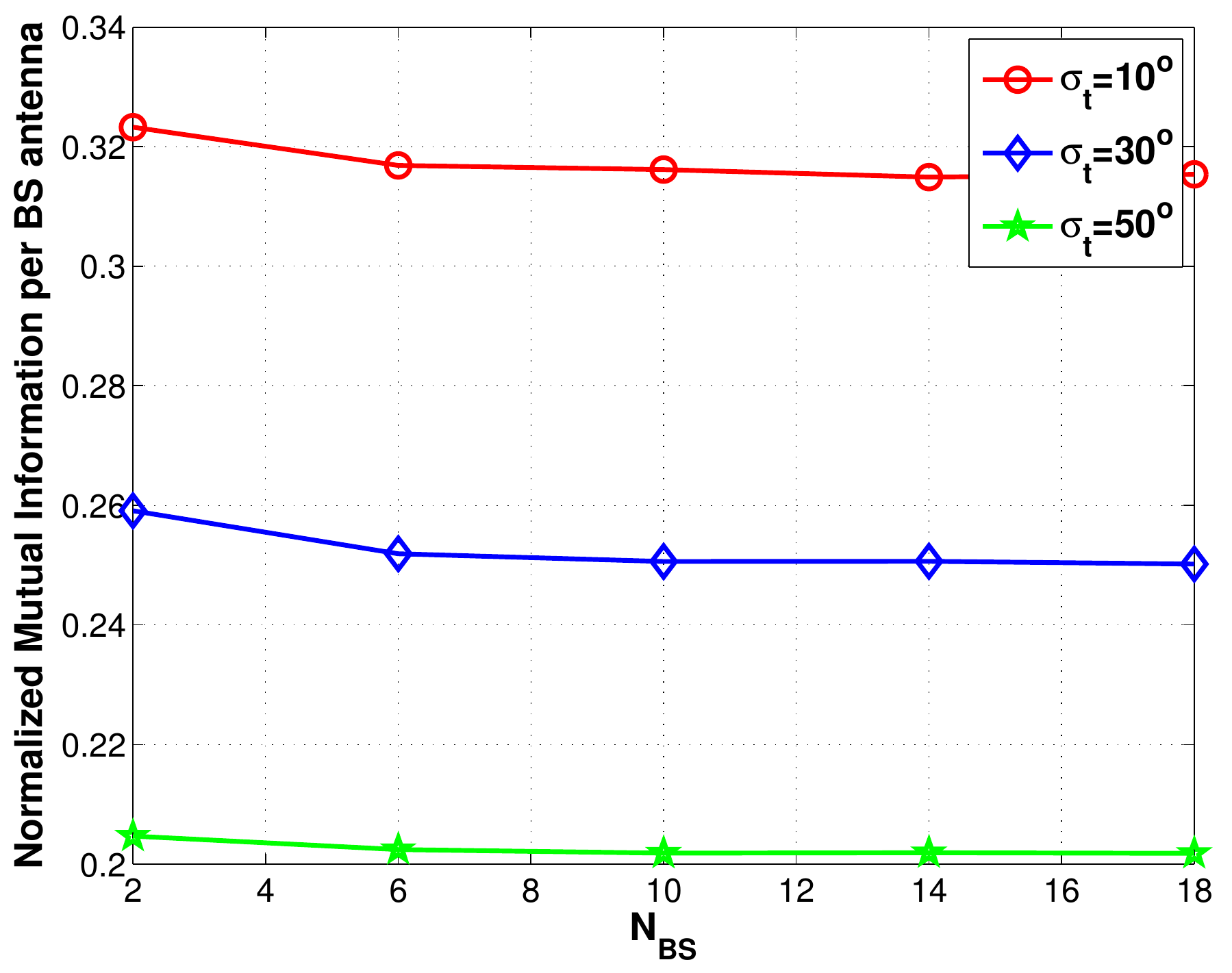}
\caption{Effect of elevation angular spread on mutual information.}
\label{sigmaMI}
\end{figure}

\begin{figure}[!t]
\centering
\includegraphics[width=2.75 in]{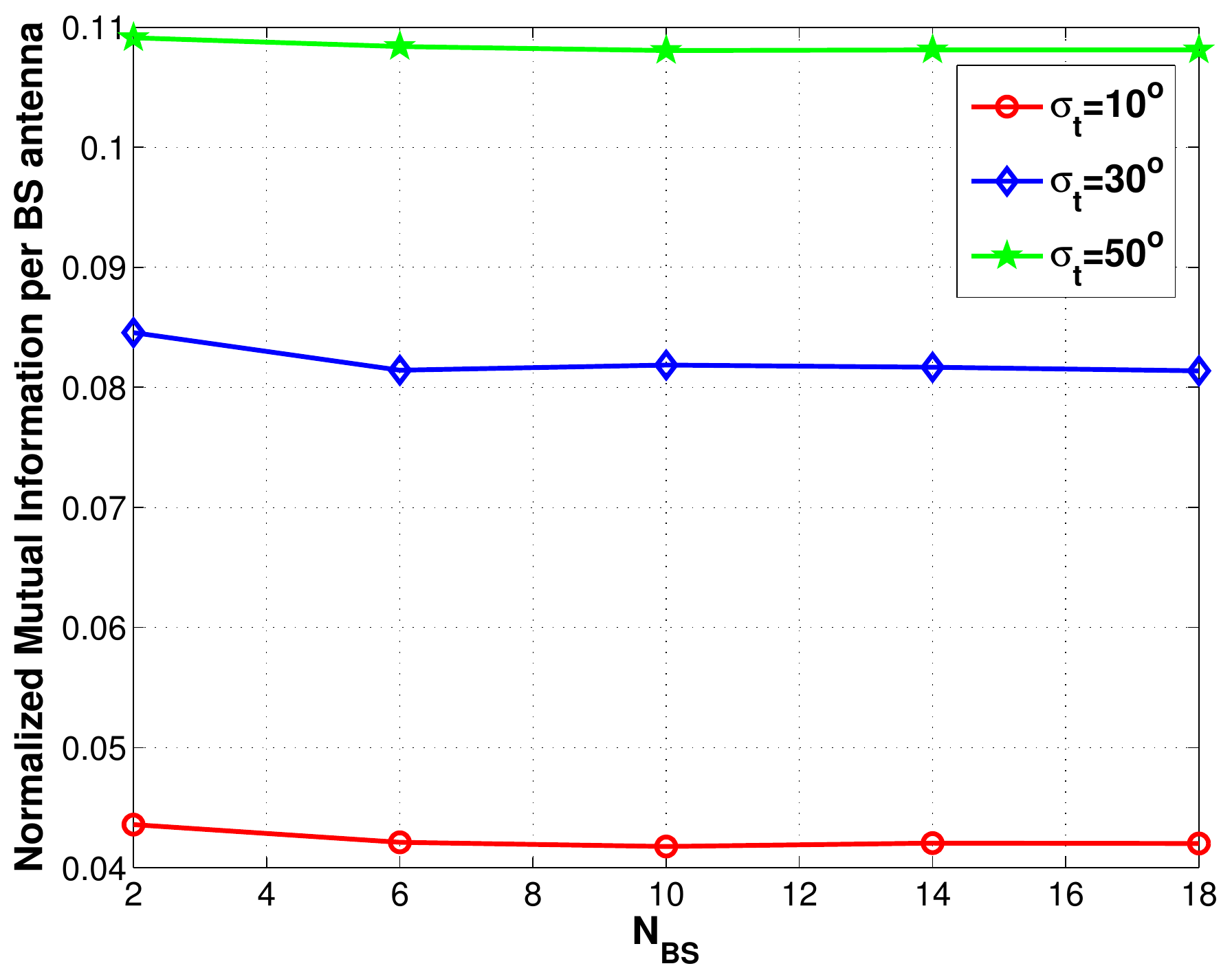}
\caption{Effect of elevation angular spread on MI for users away from the direction of antenna boresight.}
\label{sigmaMIbad}
\end{figure}

We now use the deterministic equivalent of MI to study the impact of angular spreads on the MI. It has been observed in the past that higher azimuth spread can cause the adjacent antennas to appear close to uncorrelated. This has been confirmed in the results shown in Fig. \ref{kappa} and Fig. \ref{kappaMI}. An increase in $\kappa$ corresponds to a decrease in the spread which causes the correlation to increase. A decrease in the channel MI is then evident. Only the normalized theoretical MI in (\ref{V}) for the nonparametric model has been plotted for $N=40$ to minimize the pinhole effect. The number of antennas at the BS and the MS are increased at the same rate as per Telatar's  finding that the MI scales with min($N_{BS},N_{MS}$) \cite{Telatar}. Since the correlation function is not insensitive to elevation, so in the context of 3D channels it is important to analyze the behaviour of MI with changes in angular parameters in the elevation. The correlation still decreases with an increase in the elevation angular spread as shown in Fig. \ref{sigma}. However the result for MI reveals an interesting interplay between the Tx power and the spatial correlation which is a consequence of the incorporation of the antenna pattern in our model. Note that the value of spatial correlation at 0 antenna spacing is nothing but the average Tx power of the MIMO system. It can be seen from Fig. \ref{sigma} that an increase in the elevation spread can undoubtedly cause the correlation to decrease. However, for users in good conditions with $\theta_{0} \approx \theta_{tilt}$, i.e. for users within the  direction of the antenna boresight, the incorporation of the antenna pattern into our channel model causes the Tx power to reduce with the increase in the spread as seen through the values of $\rho_{t}$ at $d_{t}/\lambda =0$. This results in an overall decrease in the MI as shown in Fig. \ref{sigmaMI}. This effect has not been observed in previous works that deal with the elevation because they generally do not consider antenna patterns in their results for SCFs and MI. It is important to note here that the incorporation of the antenna pattern would not change the expected behaviour of the increase in MI with the increase in the spread for users in bad conditions, i.e. where the user is located far away from the elevation angle of the antenna boresight. In that case, a higher spread would actually benefit this  user as it will manage to receive the signal from some propagation paths with significant energy and will benefit from the higher spread. This has been illustrated in Fig. \ref{sigmaMIbad} for the user receiving energy from the mean AoD $\theta_{0}=130^{o}$ while $\theta_{tilt}$ is still equal to $95^{o}$. 

\begin{figure}[!t]
\centering
\includegraphics[width=2.75 in]{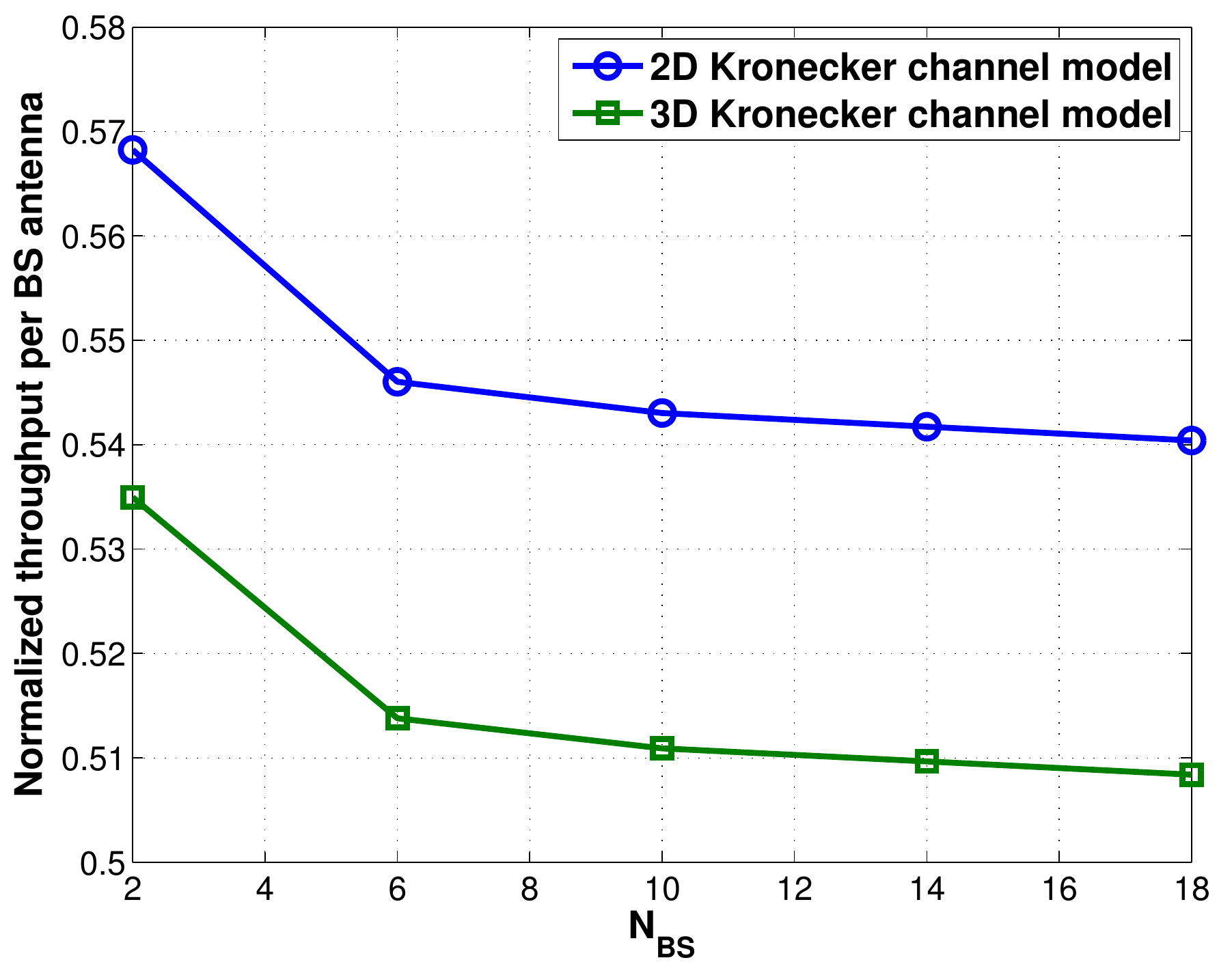}
\caption{Comparison of the MI of the 2D and 3D Kronecker channel models.}
\label{without_ele_mi}
\end{figure}

Finally we use the theoretical equivalent of MI in (\ref{V}) to compare the MI of the 2D and 3D Kronecker channel models. The former is obtained  by using the theoretical solution of (\ref{tcorr_2D}) to obtain the entries of $[\textbf{R}_{BS}]_{s,s'}=\rho_{t}(s-s')$. Similarly, the entries of $[\textbf{R}_{MS}]_{u,u'}$ are obtained by getting the theoretical solution of $\rho_{r}(u-u')=\mathbb{E}\left[g_{r,H}(\varphi)\exp\left(i\frac{2\pi}{\lambda}d_{r}(u-u')\sin\varphi\right)\right]$, after assuming $\vartheta=\pi/2$ and $g_{r,V}(\vartheta)=0$ $\rm{dB}$ in (\ref{Rx}). The result is provided in Fig. \ref{without_ele_mi} and confirms that incorporating the elevation affects the estimated values of the MI to a considerable degree. The lower MI for the 3D channel model is explained by the lower amount of transmitted power in the 3D case as seen through the value of $\rho_{t}$ at $d_{t}/\lambda =0$ in Fig. \ref{without_ele}. This phenomenon is a result of incorporating the antenna pattern in the 3D model and is observed for users located in the direction of the antenna boresight as explained earlier in the discussion on the effect of angular spreads on the MI. 


\subsection{Mutual Information Analysis of a Multi-User System}
A more robust system to channel correlation is the multiple user MIMO system, wherein instead of using multiple antennas for a single receiver, multiple users are served simultaneously. An obvious disadvantage of this system is the inter-user interference which necessitates the use of an intelligent precoding scheme at the transmitter. Regularized zero-forcing (RZF) is one of the state-of-the-art schemes that we will employ to mitigate inter-user interference in our analysis of multi-user system. The RZF precoding matrix at the BS is given by,
\begin{equation}
\label{precoding}
\textbf{G}=\sqrt{\beta} \left(\textbf{H}\textbf{H}^{H}+ \zeta N_{BS} \textbf{I}_{BS}  \right)^{-1}\textbf{H},
\end{equation}
where $\zeta$ is a strictly positive regularization parameter, $\beta$ is a scaling parameter such that $tr(\textbf{G}^{H}\textbf{G})=P,$ where $P$ is the total available Tx power and $K$ is the number of non co-operating users such that $N_{BS}\geq$ K to avoid user scheduling. The channel vector for $k^{th}$ user $\textbf{h}_{k} \in \mathbb{C}^{N_{BS}\times 1}$ is given by,
\begin{equation}
\begin{aligned}
\textbf{h}_{k}=\sqrt{\varrho_{k}} \textbf{R}_{BS,k}^{\frac{1}{2}}\textbf{z}_{k},
\end{aligned}
\end{equation}
where $\textbf{z}_{k}$ has i.i.d zero mean, unit variance complex Gaussian entries, $\textbf{R}_{BS,k}$ is the per user Tx correlation matrix and,
\begin{equation}
\label{pathloss}
\varrho_{k}=\frac{P_{Tx} \times \text{PL}_{k} \times \text{AG} \times \text{SF}}{\sigma^{2}}, 
\end{equation}
where $\text{PL}_{k}$ is the path loss experience by user $k$, SF is the shadow fading, $P_{Tx}$ is the transmitted power and \text{AG} is the antenna gain. $\text{PL}_{k}$ is computed using the path loss model proposed for Urban Macro (UMa) scenario in \cite{ITU}, SF=6 $\rm{dB}$, antenna gain=17 $\rm{dBi}$ and $\sigma^{2}$=$1.13 \times 10^{-13} W$ \cite{ITU}, \cite{TR37.84}. 

The BS uses linear precoding. The precoding vector for the $k^{th}$ user is $\textbf{g}_{k} \in \mathbb{C}^{N_{BS}\times 1}$ and the data symbol is $s_{k} \sim \mathcal{CN}$(0,1). Therefore the BS transmits the $N_{BS} \times 1$ signal,
\begin{equation}
\textbf{x}=\sum_{k=1}^{K} \textbf{g}_{k}s_{k}=\textbf{G}\textbf{s}.
\end{equation}
The SINR for the $k^{th}$ user is then given by, 
\begin{align}
\label{SINR}
\gamma_{k}&=\frac{\textbf{h}_{k}^{H}\textbf{g}_{k}\textbf{g}_{k}^{H}\textbf{h}_{k}}{\textbf{h}_{k}^{H}\textbf{G}\textbf{G}^{H}\textbf{h}_{k}-\textbf{h}_{k}^{H}\textbf{g}_{k}\textbf{g}_{k}^{H}\textbf{h}_{k}+1}.
\end{align}
\begin{figure}[!t]
\centering
\includegraphics[width=2.5 in]{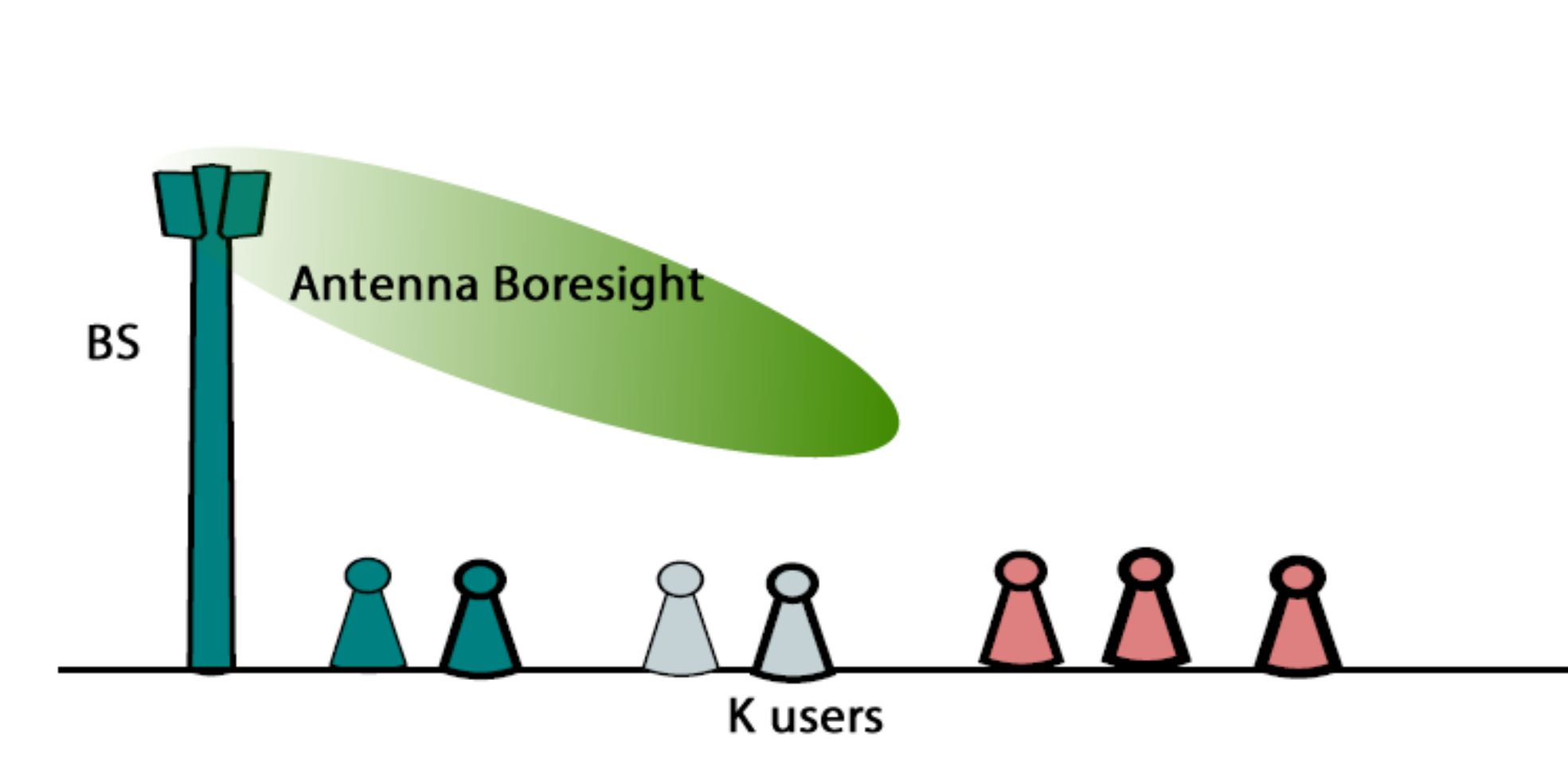}
\caption{Multi-user scenario.}
\label{tiltfig1}
\end{figure}
\begin{figure}[!t]
\centering
\includegraphics[width=2.65 in]{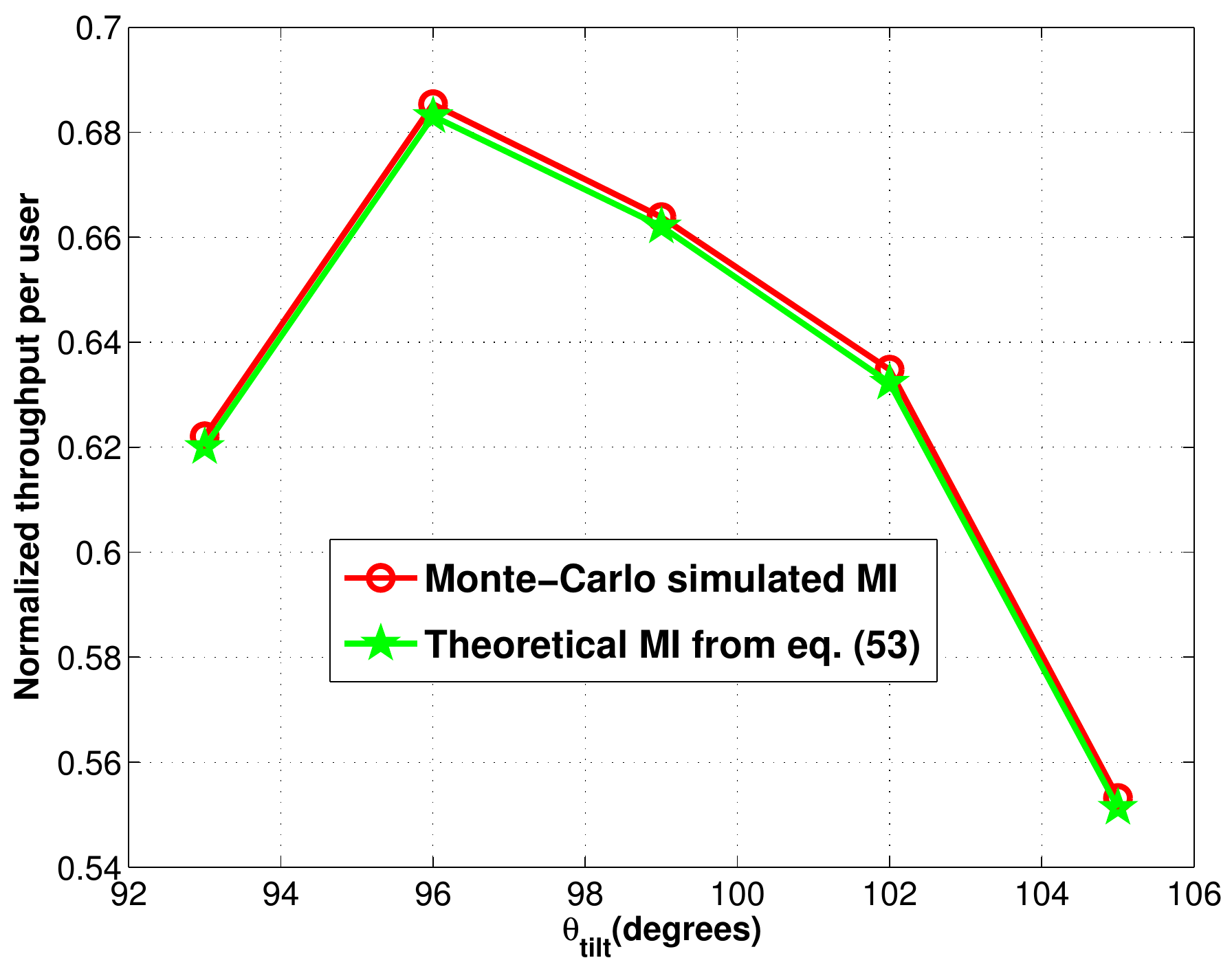}
\caption{Effect of tilt angle on mutual information in a multi-user scenario.}
\label{tiltfig}
\end{figure}

In this setting, a deterministic equivalent can be computed for the SINR, $\gamma_{k}$ for every user $k$  using tools from RMT, such that the convergence to the deterministic equivalent is almost sure as the number of antennas at BS and $K$ tend to infinity \cite{SINRdeterministic}. This deterministic equivalent was derived in great detail in \cite{SINRdeterministic} (Theorem II). The theorem says that,
\begin{equation}
\gamma_{k}-\gamma_{k}^{o} \xrightarrow[N_{BS}, K \rightarrow \infty]{a.s.} 0, 
\end{equation}
where $\gamma_{k}^{o}$ is defined in [\cite{SINRdeterministic}, equation (19) using equations (20)-(25)]. Therefore the deterministic equivalent of the MI of user $k$ in the multi-user case would be given by,
\begin{equation}
\label{detmulti}
I(\varrho_{k})=\log (1+\gamma_{k}^{o}).
\end{equation}

This convergence result holds even for a moderate number of antennas and users and facilitates the analysis of the impact of downtilt angles on the system performance in a multi-user system. We analyze the multi-user scenario with $K=40$ and $N_{BS}=60$. The scenario is illustrated in Fig. \ref{tiltfig1}. The users are randomly positioned between radii of 100m and 250m from the BS. $\varrho_{k}$ is computed using (\ref{pathloss}) for every user and the deterministic equivalent of SINR is computed based on Theorem II in  \cite{SINRdeterministic}. This helps us get a theoretical equivalent for the MI for every user. The elevation line of sight (LoS) angle with respect to the horizontal at the BS is computed for each user using $\theta_{LoS,k}=\tan^{-1} \frac{\bigtriangleup	 h}{\sqrt{(x_{k}^{2}+y_{k}^{2})}}$, where $(x_{k},y_{k})$ are the coordinates of the user $k$ and $\bigtriangleup h$ is the height difference between MS and BS.  The mean elevation AoD used in the computation of FS coefficients of PES is also equal to $\theta_{LoS,k}$. The users are located with $\theta_{LoS,k} \in [95.37^{o}, 103^{o}]$. We plot in Fig. \ref{tiltfig}, the normalized MI (Total MI of the system/$K$) of this system using the theoretical deterministic equivalent from (\ref{detmulti}) and also using the Monte-Carlo realizations of SINR in (\ref{SINR}) and see that the deterministic equivalent is quite accurate even for a moderate number of users and BS antennas. More importantly, the result shows that the performance of the users at the cell edge is most sensitive to the value of downtilt angle. The performance of the system is maximized when the antenna boresight angles at the BS are set equal to the $\theta_{LoS}$ of the user at the cell edge, i.e. $\theta_{tilt} \sim 96^{0}$. This result highlights the prospects of elevation beamforming in enhancing system performance of the future correlated MIMO systems.

\section{Conclusion}
In this paper, we characterized the spatial correlation present in a 3D MIMO channel, assuming a uniform linear array of antennas. The conventional SCFs do not take into account the effect of elevation and antenna patterns, which renders them unsuitable for the evaluation of future correlated 3D MIMO channels that are currently being outlined in the next generation of standards. We derived the proposed SCF using SHE of plane waves and properties of Legendre and associated Legendre polynomials. The final expressions for the SCF presented in \textit{Theorem 1} illustrate how this generalized function depends on the underlying arbitrary antenna patterns and angular densities through the FS coefficients of PAS and PES. Numerical results show an excellent agreement between the derived theoretical and Monte-Carlo simulated results for the spatial correlation. Furthermore, to quantify the effects of correlation on the system performance, we study the MI of the nonparametric Kronecker channel model. This model confirms the existence of pinhole channels and allows us to use the available deterministic equivalents  of the MI in the asymptotic limit for both mono-user and multi-user cases. By expressing the SCF in a closed-form as a function of channel and array parameters, the impact of azimuth and elevation angular spreads on the MI is investigated. The simulation results also provide useful insights into the impact of antenna patterns and antenna tilt angles on the achievable rates and confirm the potential of elevation beamforming to enhance the system performance.  

\appendices
\section{Proof of Theorem 1}
To derive an analytical expression for the SCF, we use the trigonometric expansion of the Legendre polynomial (\ref{legendre1}) \cite{legendre}. To this end, note that the expectation terms involving elevation angles in (\ref{prop}) can be written using the condition in (\ref{condPES}) as, 
\begin{align}
\label{FS1}
&\mathbb{E}[{P}_{2n}(\cos\theta)g_{t,V}(\theta,\theta_{tilt})]=\sum_{k=-n}^{n} \Big[ p_{n-k}p_{n+k} \int_{0}^{2\pi}\cos(2k\theta)  \nonumber \\
& \times g_{t,V}(\theta,\theta_{tilt}) p(\theta) \sin(\theta) d\theta \Big], \nonumber \\
&= \sum_{k=-n}^{n}p_{n-k}p_{n+k} \Big[\frac{1}{2}\int_{0}^{2\pi}\sin((2k+1)\theta)\text{PES}_{t}(\theta,\theta_{tilt}) d\theta \nonumber \\
&- \frac{1}{2}\int_{0}^{2\pi}\sin((2k-1)\theta)\text{PES}_{t}(\theta,\theta_{tilt})d\theta \Big].  
\end{align}
\begin{align}
\label{FS2}
&\mathbb{E}[\bar{P}_{2n}^{2m} (\cos(\theta))g_{t,V}(\theta,\theta_{tilt})]=\sum_{k=0}^{n} \Big[ c_{2n,2k}^{2m} \int_{0}^{2\pi}\cos(2k\theta) \nonumber \\
& \times g_{t,V}(\theta,\theta_{tilt})  p(\theta) \sin(\theta) d\theta \Big], \nonumber \\
&= \sum_{k=0}^{n}c_{2n,2k}^{2m} \Big[\frac{1}{2} \int_{0}^{2\pi}\sin((2k+1)\theta)\text{PES}_{t}(\theta,\theta_{tilt}) d\theta \nonumber \\
&- \frac{1}{2}\int_{0}^{2\pi}\sin((2k-1)\theta)\text{PES}_{t}(\theta,\theta_{tilt})d\theta \Big].
\end{align}
\begin{align}
\label{FS3}
&\mathbb{E}[\bar{P}_{2n-1}^{2m-1} (\cos(\theta)) g_{t,V}(\theta,\theta_{tilt})]=\sum_{k=1}^{n} \Big[d_{2n-1,2k-1}^{2m-1}  \nonumber \\
& \times \int_{0}^{2\pi}\sin((2k-1)\theta) g_{t,V}(\theta,\theta_{tilt})p(\theta)\sin(\theta) d\theta \Big],  \nonumber \\
&=\sum_{k=1}^{n}d_{2n-1,2k-1}^{2m-1} \Big[\frac{1}{2}\int_{0}^{2\pi}\cos((2k-2)\theta)\text{PES}_{t}(\theta,\theta_{tilt}) d\theta \nonumber \\
&- \frac{1}{2}\int_{0}^{2\pi}\cos((2k)\theta)\text{PES}_{t}(\theta,\theta_{tilt})d\theta \Big].
\end{align}
As explained in section II-C, the limits in (\ref{FS1})-(\ref{FS3}) have been set to (0, 2$\pi$) as $f_{\theta}(\theta)$ is 0 outside (0, $\pi$). Doing this, we have expressed the expectations involving the elevation angles as a linear combination of the scaled FS coefficients of PES. For the expectations involving azimuth angles,
\begin{align}
\label{FS4}
&\mathbb{E}[\cos(2m\phi) g_{t,H}(\phi)]=\int_{-\pi}^{\pi}\cos(2m\phi)g_{t,H}(\phi) f(\phi) d\phi, \nonumber \\
&= \int_{-\pi}^{\pi}\cos(2m\phi)\text{PAS}_{t}(\phi) d\phi. 
\end{align}
\begin{align}
\label{FS5}
&\mathbb{E}[\sin((2m-1)\phi)g_{t,H}(\phi)]=\int_{-\pi}^{\pi}\sin((2m-1)\phi)g_{t,H}(\phi) \nonumber \\
& \times f(\phi)d\phi = \int_{-\pi}^{\pi}\sin((2m-1)\phi)\text{PAS}_{t}(\phi)d\phi.
\end{align}
\normalsize
They express the expectations involving azimuth angles as a linear combination of scaled FS coefficients of PAS. Defining the FS coefficients of PAS and PES as,
\begin{align}
\label{a1}
a_{\phi}(m)&= \frac{1}{\pi}\int_{-\pi}^{\pi}\text{PAS}_{t}(\phi)\cos(m\phi) d\phi, \\
\label{b1}
b_{\phi}(m)&= \frac{1}{\pi}\int_{-\pi}^{\pi}\text{PAS}_{t}(\phi)\sin(m\phi) d\phi,  \\
\label{a2}
a_{\theta}(k)&=\frac{1}{\pi}\int_{0}^{2\pi}\text{PES}_{t}(\theta,\theta_{tilt})\cos(k\theta) d\theta ,  \\
\label{b2}
b_{\theta}(k)&= \frac{1}{\pi}\int_{0}^{2\pi}\text{PES}_{t}(\theta,\theta_{tilt})\sin(k\theta) d\theta, 
\end{align}
one can immediately see the relationship between the terms involving expectations and these coefficients. Plugging (\ref{a1})-(\ref{b2}) in (\ref{FS1})-(\ref{FS5}) and using the resulting expressions in (\ref{prop}) would result in a closed-form expression for $\rho_{t}(s-s')$ yielding (\ref{tcorr}). A similar development would yield (\ref{rcorr}) by replacing AoD with AoA and $g_{t}$ replaced with $g_{r}$. This completes the proof of Theorem 1.

\section{Fourier Series Coefficients of PAS and PES}
\subsection{FS Coefficients of PES at BS}
The FS coefficients are computed assuming $G_{p,max}=0$ $\rm{dB}$. The value of $G_{p,max}=17$ $\rm{dBi}$ will be incorporated in the simulations as a scaling factor (Antenna Gain).
\begin{align*}
\small
&a_{\theta,1}(m)=\frac{A}{\pi} \int_{\theta_{0}}^{\pi}\exp\left(-\frac{\sqrt{2}(\theta-\theta_{0})}{\sigma}\right)  \\
& \times \exp\left(-1.2\left(\frac{\theta-\theta_{tilt}}{\theta_{3dB}} \right)^{2}\log(10)\right) \cos(m \theta) d\theta, \\
&=\frac{A}{2\sqrt{a\pi}} \Re \Big[ \exp\left(-c+\frac{2\pi a j m+b^{2}+2bjm-m^{2}}{4a}\right) \\
& \times \left(erf \left(\frac{2a\frac{\pi}{2}+b+j m}{2\sqrt{a}} \right) -erf \left(\frac{2a(\frac{\pi}{2}-\theta_{0})+b+jm}{2\sqrt{a}} \right) \right) \Big],  
\end{align*}
where $a=\frac{1.2\log(10)}{\theta_{3dB}^{2}}$, $b=\frac{-2.4(\frac{\pi}{2}-\theta_{tilt})\log(10)}{\theta_{3dB}^{2}}+\frac{\sqrt{2}}{\sigma}$ and $c=\frac{1.2(\frac{\pi}{2}-\theta_{tilt})^{2}\log(10)}{\theta_{3dB}^{2}}-\frac{\sqrt{2}(\frac{\pi}{2}-\theta_{0})}{\sigma}$ and $erf$ is error function. 
\begin{align*}
\small
&a_{\theta,2}(m)=\frac{A}{\pi} \int_{0}^{\theta_{0}}\exp\left(-\frac{\sqrt{2}(\theta_{0}-\theta)}{\sigma}\right)  \\
& \times \exp\left(-1.2\left(\frac{\theta-\theta_{tilt}}{\theta_{3dB}} \right)^{2}\log(10)\right) \cos(m \theta) d\theta, \\
&=\frac{A}{2\sqrt{a\pi}} \Re \Big[ \exp\left(-c+\frac{2\pi a jm+b^{2}+2bjm-m^{2}}{4a}\right) \\
& \times \left(erf\left(\frac{2a(\frac{\pi}{2}-\theta_{0})+b+jm}{2\sqrt{a}} \right) -erf\left(\frac{-2a\frac{\pi}{2}+b+jm}{2\sqrt{a}} \right) \right) \Big], 
\end{align*}
where $a=\frac{1.2\log(10)}{\theta_{3dB}^{2}}$, $b=\frac{-2.4(\frac{\pi}{2}-\theta_{tilt})\log(10)}{\theta_{3dB}^{2}}-\frac{\sqrt{2}}{\sigma}$ and $c=\frac{1.2(\frac{\pi}{2}-\theta_{tilt})^{2}\log(10)}{\theta_{3dB}^{2}}+\frac{\sqrt{2}(\frac{\pi}{2}-\theta_{0})}{\sigma}$. 

Also $b_{\theta,1}(m)$ and $b_{\theta,2}(m)$ have the same expressions as $a_{\theta,1}(m)$ and $a_{\theta,2}(m)$ respectively with only $\Re$ replaced with $\Im$. The FS coefficients are,
\begin{align}
a_{\theta}(m)&=a_{\theta,1}(m)+a_{\theta,2}(m), \\
b_{\theta}(m)&=b_{\theta,1}(m)+b_{\theta,2}(m). 
\end{align}
\subsection{FS Coefficients of PES at MS}
Note that $g_{r}(\vartheta)=1$.
\begin{align*}
&a_{\vartheta}(m)=\frac{A}{\pi} \int_{0}^{\pi}\exp\left(-\frac{\sqrt{2}|\vartheta-\vartheta_{0}|}{\sigma}\right) \cos\left(m \theta\right) d\theta, \\
&=\frac{A \sigma^{2}}{\pi (2+m^{2}\sigma^{2})} \Big[\frac{2\sqrt{2}}{\sigma} \cos\left(m \theta_{0}\right) - \frac{\sqrt{2}}{\sigma} \exp\left(\frac{-\pi}{\sqrt{2}\sigma}\right) \\
& \times \left( \exp \left(\frac{\sqrt{2}(\frac{\pi}{2}-\theta_{0})}{\sigma}\right) + (-1)^{m} \exp \left(\frac{-\sqrt{2}(\frac{\pi}{2}-\theta_{0})}{\sigma}\right)\right)     \Big],  \\
&b_{\vartheta}(m)=\frac{A}{\pi} \int_{0}^{\pi}\exp\left(-\frac{\sqrt{2}|\vartheta-\vartheta_{0}|}{\sigma}\right) \sin\left(m \theta\right) d\theta, \\
&=\frac{A \sigma^{2}}{\pi (2+m^{2}\sigma^{2})} \Big[\frac{2\sqrt{2}}{\sigma} \sin\left(m \theta_{0}\right) + m \exp\left(\frac{-\pi}{\sqrt{2}\sigma}\right) \\
& \times \left( \exp \left(\frac{\sqrt{2}(\frac{\pi}{2}-\theta_{0})}{\sigma}\right) - (-1)^{m} \exp \left(\frac{-\sqrt{2}(\frac{\pi}{2}-\theta_{0})}{\sigma}\right)\right)     \Big]. 
\end{align*}

\subsection{FS Coefficients of PAS}
Since the antennas are considered omnidirectional in the azimuth, so $g_{t,H}(\phi)$ and $g_{r,H}(\varphi)$=1. The VM distribution can be expressed as a series of Bessel functions as,
\begin{align}
&\text{PAS}(\phi)=\frac{1}{2\pi} \left(1+\frac{2}{I_{0}(\kappa)}\sum_{j=1}^{\infty}I_{j}(\kappa)\cos(j(\phi - \mu)) \right), \\
&=\frac{1}{2\pi}+\frac{1}{\pi I_{0}(\kappa)}\sum_{j=1}^{\infty}I_{j}(\kappa)[\cos(j\phi)\cos(j\mu) + \sin(j\phi)\sin(j\mu)] \\
&a_{\phi}(m)=\frac{1}{\pi I_{0}(\kappa)} I_{m}(\kappa)\cos(m \mu), \\
&b_{\phi}(m)=\frac{1}{\pi I_{0}(\kappa)} I_{m}(\kappa)\sin(m \mu).
\end{align}

\bibliographystyle{IEEEtran}
\bibliography{bib}

\begin{IEEEbiography}[{\includegraphics[width=1 in, height=1.25 in,clip,keepaspectratio ]{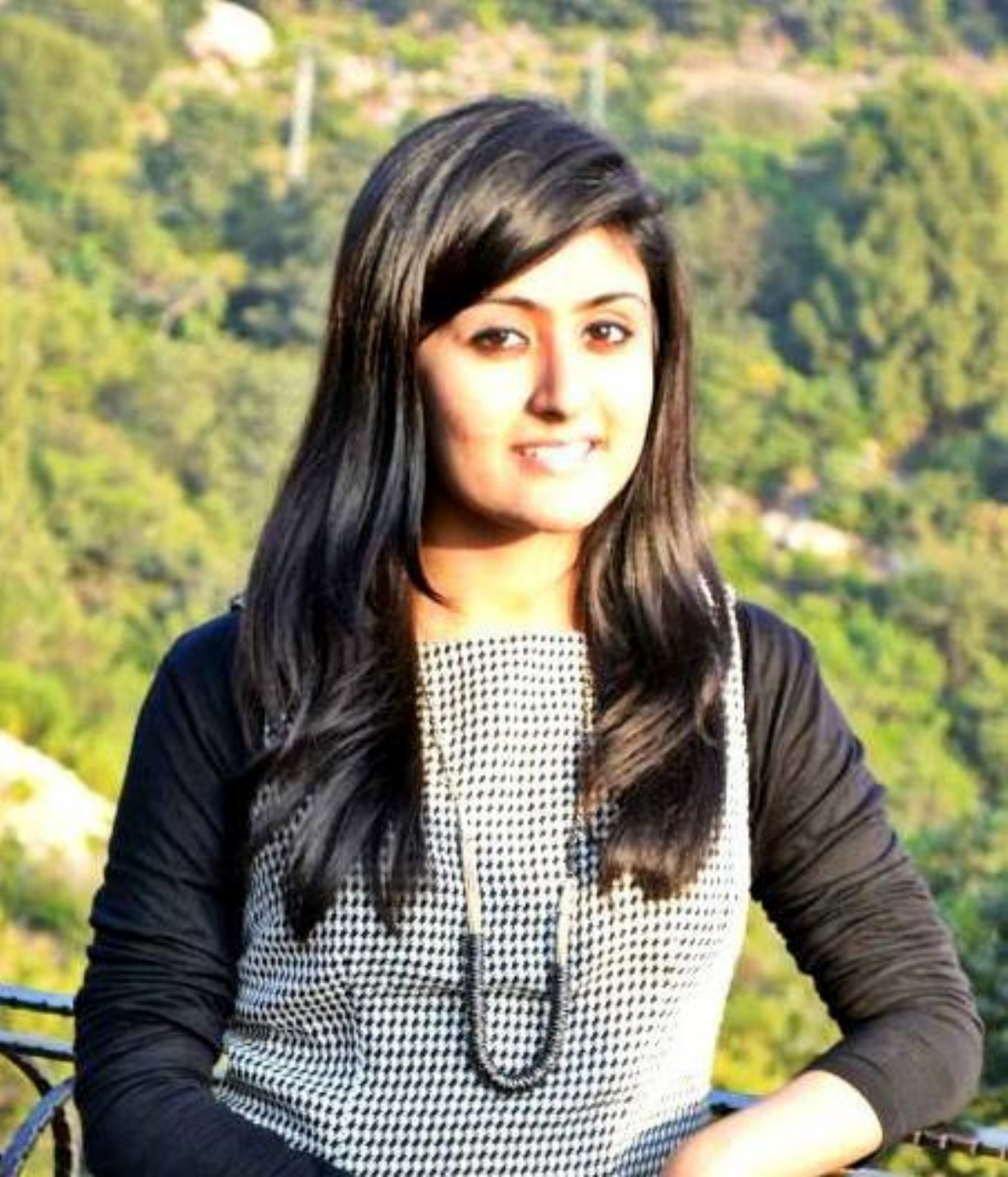}}]{Qurrat-Ul-Ain Nadeem}
Qurrat-Ul-Ain Nadeem was born in Lahore, Pakistan. She received her B.S. degree in electrical engineering from the Lahore University of Management Sciences (LUMS), Pakistan in 2013. She joined King Abdullah University of Science and Technology (KAUST), Thuwal, Makkah Province, Saudi Arabia in August 2013, where she is currently a M.S./Ph.D. student. Her research interests include channel modeling and performance analysis of wireless communications systems. 
\end{IEEEbiography}

\begin{IEEEbiographynophoto}{Abla Kammoun}
Abla Kammoun was born in Sfax, Tunisia. She received the engineering degree in signal and systems from the Tunisia Polytechnic School, La Marsa, and the Master's degree and the Ph.D. degree in digital communications from Telecom Paris Tech [then Ecole Nationale Sup{\'e}rieure des T{\'e}l{\'e}communications (ENST)]. From June 2010 to April 2012, she has been a Postdoctoral Researcher in the TSI Department, Telecom Paris Tech. Then she has been at Sup{\'e}lec at the Large Networks and Systems Group (LANEAS) until December 2013. Currently, she is  research scientist at KAUST university. Her research interests include performance analysis, random matrix theory, and semi-blind channel estimation.
\end{IEEEbiographynophoto}


\begin{IEEEbiographynophoto}{M{\'e}rouane Debbah}
M{\'e}rouane Debbah entered the Ecole Normale Sup{\'e}rieure de Cachan (France) in 1996 where he received his M.Sc and Ph.D. degrees respectively. He worked for Motorola Labs (Saclay, France) from 1999-2002 and the Vienna Research Center for Telecommunications (Vienna, Austria) until 2003. From 2003 to 2007, he joined the Mobile Communications department of the Institut Eurecom (Sophia Antipolis, France) as an Assistant Professor. Since 2007, he is a Full Professor at Sup{\'e}lec (Gif-sur-Yvette, France). From 2007 to 2014, he was director of the Alcatel-Lucent Chair on Flexible Radio. Since 2014, he is Vice-President of the Huawei France R\&D center and director of the Mathematical and Algorithmic Sciences Lab. His research interests lie in fundamental mathematics, algorithms, complex systems analysis and optimization, statistics, information \& communication sciences research. He is an Associate Editor in Chief of the journal Random Matrix: Theory and Applications and was an associate and senior area editor for IEEE Transactions on Signal Processing respectively in 2011-2013 and 2013-2014. M{\'e}rouane Debbah is a recipient of the ERC grant MORE (Advanced Mathematical Tools for Complex Network Engineering). He is a IEEE Fellow, a WWRF Fellow and a member of the academic senate of Paris-Saclay. He is the recipient of the Mario Boella award in 2005, the 2007 IEEE GLOBECOM best paper award, the Wi-Opt 2009 best paper award, the 2010 Newcom++ best paper award, the WUN CogCom Best Paper 2012 and 2013 Award, the 2014 WCNC best paper award as well as the Valuetools 2007, Valuetools 2008, CrownCom2009 , Valuetools 2012 and SAM 2014 best student paper awards. In 2011, he received the IEEE Glavieux Prize Award and in 2012, the Qualcomm Innovation Prize Award.
\end{IEEEbiographynophoto}

\begin{IEEEbiography}[{\includegraphics[width=1 in, height=1.25 in,clip,keepaspectratio ]{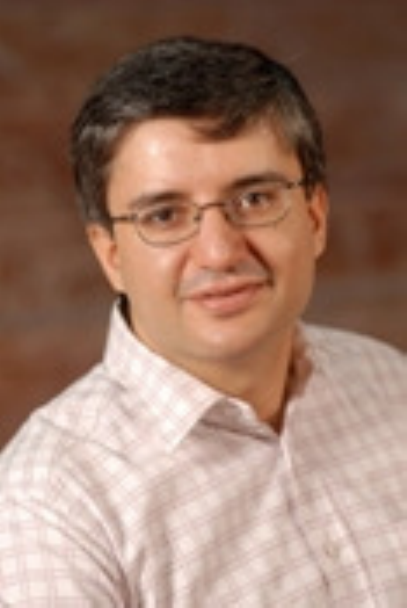}}]{Mohamed-Slim Alouini} (S'94, M'98, SM'03, F’09) Mohamed-Slim Alouini was born in Tunis, Tunisia. He received the Ph.D. degree in Electrical Engineering from the California Institute of Technology (Caltech), Pasadena, CA, USA, in 1998. He served as a faculty member in the University of Minnesota, Minneapolis, MN, USA, then in the Texas A\&M University at Qatar, Education City, Doha, Qatar before joining King Abdullah University of Science and Technology (KAUST), Thuwal, Makkah Province, Saudi Arabia as a Professor of Electrical Engineering in 2009. His current research interests include the modeling, design, and performance analysis of wireless communication systems.
\end{IEEEbiography}

\end{document}